\documentclass[aps,pre,reprint]{revtex4-1}

\usepackage{microtype}
\usepackage[pdfborder={0 0 0}]{hyperref} 
\usepackage{mathtools}
\usepackage{amstext}
\usepackage{amsmath}
\usepackage{amssymb}
\usepackage{amsfonts}
\usepackage{graphicx}
\usepackage{color}
\usepackage[normalem]{ulem} 

\renewcommand{\vec}[1]{\boldsymbol{#1}}  
\newcommand*{\diff}{\mathrm{d}}
\newcommand{\del}{\partial}
\newcommand{\Lap}{\Delta}
\let\sphi\phi 
\let\phi\varphi 

\let\epsilon\varepsilon

\definecolor{Red}{rgb}{0.9,0.0,0.1}
\definecolor{Lila}{rgb}{0.7,0.0,0.9}
\definecolor{Darkblue}{rgb}{0.22,0.33,0.64}
\definecolor{Darkgray}{rgb}{0.4,0.4,0.4}
\definecolor{Blue}{rgb}{0.1,0.0,0.9}
\newcommand{\Cr}[1]{#1}

\begin{document}

\title{The secondary buckling transition: wrinkling of buckled spherical shells}
\author{Sebastian Knoche} 
\affiliation{Department of Physics, Technische Universit\"{a}t Dortmund, 44221 Dortmund, Germany}

\author{Jan Kierfeld} 
\email{jan.kierfeld@tu-dortmund.de}
\affiliation{Department of Physics, Technische Universit\"{a}t Dortmund, 44221
  Dortmund, Germany}

\begin{abstract} 
  We theoretically explain the complete sequence of shapes of deflated spherical shells. Decreasing the volume, the shell remains spherical initially, then undergoes the classical buckling instability, where an axisymmetric dimple appears, and, finally, loses its axisymmetry by wrinkles developing in the vicinity of the dimple edge in a secondary buckling transition. We describe the first axisymmetric buckling transition by numerical integration of the complete set of shape equations and an approximate analytic model due to Pogorelov. In the buckled shape, both approaches exhibit a locally compressive hoop stress in a region where experiments and simulations show the development of polygonal wrinkles, along the dimple edge. In a simplified model based on the stability equations of shallow shells, a critical value for the compressive hoop stress is derived, for which the compressed circumferential fibres will buckle out of their circular shape in order to release the compression. By applying this wrinkling criterion to the solutions of the axisymmetric models, we can calculate the critical volume for the secondary buckling transition. Using the Pogorelov approach, we also obtain an analytical expression for the critical volume at the secondary buckling transition: The critical volume difference scales linearly with the bending stiffness, whereas the critical volume reduction at the classical axisymmetric buckling transition scales with the square root of the bending stiffness. These results are confirmed by another stability analysis in the framework of Donnel, Mushtari and Vlasov (DMV) shell theory, and by numerical simulations available in the literature.
\end{abstract}
%

\maketitle

\section{Introduction}

\begin{figure}[b]
  \centerline{\includegraphics[width=80mm]{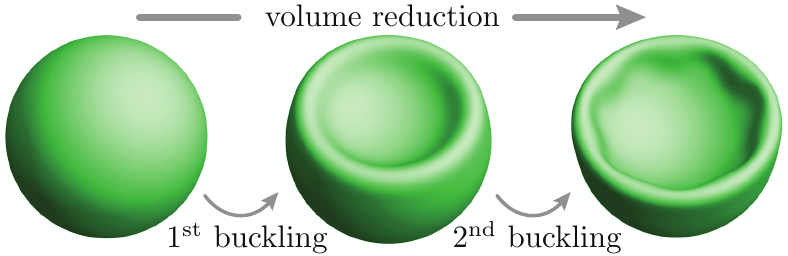}}
  \caption{(Colour online) Typical course of a capsule deflation.}
\label{fig:3Dplots}
\end{figure}

When spherical shells, such as sports and toy balls or microcapsules, are deflated, they always go through the same sequence of shapes, see fig.\ \ref{fig:3Dplots}. At small deflation, the capsule remains spherical. Upon reducing the volume, an axisymmetric dimple forms in an abrupt transition. For sufficiently thin shells, this dimple finally loses its axisymmetry upon further volume reduction, resulting in a polygonally buckled shape. This is shown by daily life experience, microcapsule experiments \cite{Quilliet2008,Datta2010,Datta2012}, and computer simulations \cite{Quilliet2008,Vliegenthart2011,Quilliet2012} based on triangulated surfaces (e.g.\ with the surface evolver) or finite element methods \cite{Vaziri2008,Vaziri2009,Vella2011}. This generic behaviour also applies to other types of deformation, for example when a dimple is formed by indenting the capsule with a point force \cite{Vaziri2008,Vaziri2009,Pauchard1998}, when the capsule is pressed between rigid plates \cite{Vaziri2009} or when the capsule adheres to a substrate \cite{Komura2005}.

The first buckling transition, where an axisymmetric dimple forms, is a classical problem in continuum mechanics, which is very well understood. Linear shell theory can be  used to calculate the onset of instability of the spherical shape, see references \cite{Landau1986,Ventsel2001}. Furthermore, nonlinear shell theory has been used to investigate the postbuckling behaviour \cite{Koiter1969}, which revealed that the buckled shape is unstable with respect to further volume reduction if the pressure is controlled. Numerical analyses of nonlinear shape equations have been used to obtain bifurcation diagrams for the axisymmetric deformation behaviour \cite{Bauer1970,Knoche2011}. In this paper, we will approach this classical axisymmetric buckling using  two different models for axisymmetric deformations. The first model is based on nonlinear shell theory, which leads to a complete set of shape equations which are to be solved numerically \cite{Knoche2011}. The second model is an  approximate analytical model which has  been proposed by Pogorelov \cite{Pogorelov1988}.

In contrast, the secondary buckling transition, where the dimple loses its axisymmetry, has been merely observed experimentally or in computer experiments so far. A theoretical approach which explains the mechanism underlying this instability and which can predict the corresponding critical volume for the secondary buckling transition is still lacking. Here we will offer an explanation within continuum elasticity theory. This also demonstrates  that polygonal capsule shapes can also occur in the absence of any discretisation effects. For crystalline elastic capsules, defects in the triangulation gives rise to additional faceting effects upon deflation \cite{Yong2013}.

Analysing the results of the axisymmetric models, we get a hint on the physical mechanism of the secondary buckling transition: we observe a region of compressive hoop stress, which is located in the inner neighbourhood of the dimple edge, just in the place where the secondary buckling occurs in experiments and simulations. In order to release the compressive stress, the circumferential fibres buckle out of their circular shape if the hoop stress reaches a critical value; quite comparable to the Euler buckling of straight rods \cite{Landau1986}. A quantitative investigation of the secondary buckling transition therefore consists of two steps: Firstly, quantifying the stress distribution in the axisymmetric buckled configuration, and secondly, finding the critical compressive stress at which the axisymmetric configuration loses its stability.

The first task is readily accomplished by the two axisymmetric models, the approach by nonlinear shape equations or Pogorelov's approximate analytical model. For the second task, we will develop a simplified model which captures the essential features of the geometry and stress distribution of the axisymmetric buckled shape, and derive a critical compressive stress by using the stability equations of shallow shells \cite{Ventsel2001}.

We verify our results by a second approach in which we conduct a stability analysis of the full shape, where the exact geometry and shape distribution is taken into account. The framework to do this is the nonlinear DMV (Donnel, Mushtari and Vlasov) theory of shells \cite{Niordson1985,Ventsel2001}.

The resulting critical volumes for the primary and secondary buckling transition are presented in a phase or stability diagram in fig.\ \ref{fig:phase_diag}, from which the state of a deflated capsule with given bending stiffness and volume difference can be read off and which is the main result of the paper. A short account of these results focusing on the secondary buckling mechanism and the parameter dependencies of the critical buckling volumes has been given elsewhere \cite{Knoche2014}.

\section{Axisymmetric primary buckling of shells}

\subsection{Nonlinear shape equations}
\label{sec:nonlinear_shape_equations}

\begin{figure}[t]
  \centerline{\includegraphics[width=80mm]{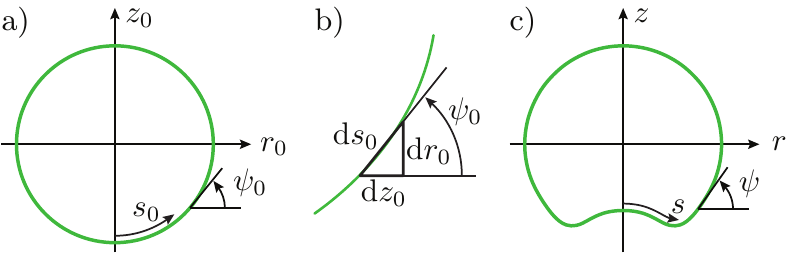}}
  \caption{(Colour online) Geometry of the axisymmetric midsurface. a) Undeformed shape (always with index ``0''), b) definition of the slope angle $\psi_0$, c) deformed shape.}
\label{fig:midsurface}
\end{figure}

Figure \ref{fig:midsurface} shows the parametrisation of the capsule midsurface. The undeformed shape is parametrised in cylindrical polar coordinates by the functions $r_0(s_0)$ and $z_0(s_0)$, where $s_0 \in [0, L_0]$ is the arc length. The slope angle $\psi_0$, defined by the two relations
\begin{equation}
 \frac{\diff r_0}{\diff s_0} = \cos \psi_0 \quad \text{and} \quad 
 \frac{\diff z_0}{\diff s_0} = \sin \psi_0,
 \label{eq:psi_0_def}
\end{equation}
see fig.\ \ref{fig:midsurface} b), permits simple calculation of the principal curvatures of the midsurface. These are found in meridional and circumferential direction, respectively, and read
\begin{equation}
 \kappa_{s_0} = \frac{\diff \psi_0}{\diff s_0} \quad \text{and} \quad 
 \kappa_{\phi_0} = \frac{\sin \psi_0}{r_0}.
 \label{eq:kappa_def}
\end{equation} 
In our case, the spherical reference configuration is explicitly given by $r_0(s_0) = R_0 \sin(s_0/R_0)$ and $z_0(s_0) = - R_0 \cos(s_0/R_0)$, and the principal curvatures reduce to $\kappa_{s_0} = \kappa_{\phi_0} = 1/R_0$. The geometrical relations (\ref{eq:psi_0_def}) and (\ref{eq:kappa_def}) also hold for the deformed midsurface when all indices ``0'' are omitted.

Upon deformation into a different axisymmetric shape, the midsurface undergoes stretching and bending. We measure the stretches in meridional and circumferential direction by
\begin{equation}
 \lambda_s = \diff s / \diff s_0 \quad \text{and} \quad \lambda_\phi = r/r_0,
 \label{eq:lambda_def} 
\end{equation} 
respectively. The function $s(s_0)$ defined in this context determines the position $s$ at which a shell element originally located at $s_0$ can be found after deformation. The strains that correspond to these stretches are centred around $0$ and are defined as $e_s = \lambda_s - 1$ and $e_\phi = \lambda_\phi -1 $. To measure the change in curvature, the bending strains
\begin{equation}
 K_s= \lambda_s \, \kappa_s - \kappa_{s_0} \quad \text{and} \quad 
K_\phi = \lambda_\phi \, \kappa_\phi - \kappa_{\phi_0}
 \label{eq:bending_strain}
\end{equation} 
are defined \cite{Libai1998,Pozrikidis2003}.

The deformation results in an elastic energy which is stored in the membrane. This elastic energy can be calculated as the surface integral over an elastic energy density (measured per undeformed surface area), which we assume to be of the simple Hookean form
\begin{multline}
 w_S =  \frac{1}{2} \frac{EH_0}{1-\nu^2} 
  \left( e_s^2 + 2\, \nu \,e_s\, e_\phi + e_\phi^2 \right) \\
   + \frac{1}{2} E_B  \left(  K_s^2 + 2 \,\nu\, K_s \,K_\phi + K_\phi^2
   \right).
\label{eq:w_S}
\end{multline}
In this expression, $E$ is the (three-dimensional) Young modulus, $H_0$ the membrane thickness, $\nu$ is the (three-dimensional) Poisson ratio which is confined to $-1 \leq \nu \leq 1/2$, and $E_B = EH_0^3/12\big( 1 - \nu^2 \big)$ is the bending stiffness. It can be shown \cite{Libai1998} by the principle of virtual work that the meridional tension and bending moment can be derived as
\begin{equation}
 \begin{aligned}
 \tau_s &= \frac{1}{\lambda_\phi} \frac{\del w_S}{\del e_s}
   = \frac{E H_0}{1-\nu^2} \, \frac{1}{\lambda_\phi} 
  \big( e_s + \nu\, e_\phi \big), \\
    m_s &= \frac{1}{\lambda_\phi} \frac{\del w_S}{\del K_s}
   = E_B \, \frac{1}{\lambda_\phi}\big( K_s + \nu\, K_\phi \big).
\end{aligned}
\label{eq:stress-strain_2D}
\end{equation} 
from the energy density. The corresponding relations for the circumferential tension and bending moment are obtained by interchanging all indices $s$ and $\phi$ in these equations. Note that the tensions and bending moments are measured per unit length of the deformed midsurface, whereas the energy density is measured per unit area of the undeformed midsurface -- this is the reason why the factors $1/\lambda_\phi$ occur in (\ref{eq:stress-strain_2D}).

To close the problem of determining the deformed shape, we need equations of equilibrium. These read, for the tangential force, normal force and bending moment,
\begin{align}
 0 &= - \frac{\cos \psi}{r}\, \tau_\phi 
  + \frac{1}{r}\, \frac{\diff(r\,\tau_s)}{\diff s} 
   - \kappa_s\,q, \label{eq:equil1}\\
 0 &= -p + \kappa_\phi \, \tau_\phi 
   + \kappa_s \, \tau_s 
   + \frac{1}{r}\, \frac{\diff(r\,q)}{\diff s}, \label{eq:equil2}\\
 0 &=  \frac{\cos \psi}{r} \, m_\phi 
  - \frac{1}{r} \frac{\diff(r\,m_s)}{\diff s} - q.\label{eq:equil3}
\end{align}
In these equations, $q$ is the transversal shear force, and $p$ is the applied normal pressure, which can also be interpreted as a Lagrange multiplier to control the capsule volume.

For a numerical treatment, the equilibrium equations should be written as a system of first-order differential equations, which are called \emph{shape equations}. They follow from (\ref{eq:psi_0_def}), (\ref{eq:kappa_def}) and (\ref{eq:equil1}) - (\ref{eq:equil3}), see \cite{Knoche2011}. Boundary conditions must be imposed which assure that the capsule is closed and has no kinks at the poles. In ref.\ \cite{Knoche2011}, a detailed discussion of this issue is given, as well as numerical procedures that are suitable for solving the shape equations.

For the further analysis, it is convenient to introduce dimensionless quantities by measuring tensions in units of $E H_0$ and lengths in units of
$R_0$. Specifically, this results in a dimensionless bending stiffness $\tilde
E_B$ which is the inverse of the F\"oppl-von-K\'arm\'an-number
$\gamma_\text{FvK}$,
\begin{equation}
 \tilde E_B \equiv \frac{E_B}{E H_0 R_0^2}
 \quad  \text{and} \quad 
 \gamma_\text{FvK} = \frac{1}{\tilde E_B}.
\end{equation}

\subsection{Analytic Pogorelov model}

\begin{figure}[t]
  \centerline{\includegraphics[width=80mm]{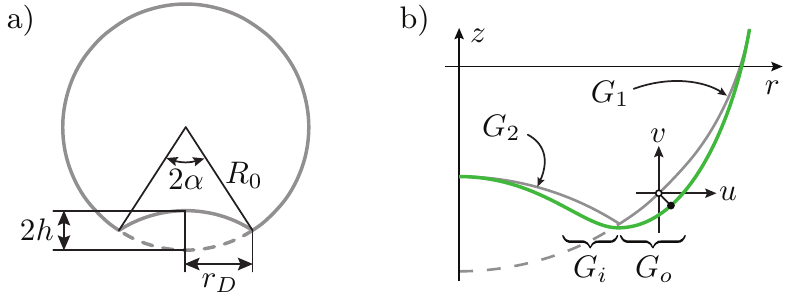}}
  \caption{(Colour online) Midsurface geometry in the analytic model. a) The isometric deformation, where a spherical cap of radius $r_D$ and depth $h$ is mirror inverted. b) The final shape (green line) differs from the isometric shape by small displacements $(u, v)$.}
\label{fig:midsurface_pog}
\end{figure}

The second approach to axisymmetric buckled shapes is based on a model of Pogorelov \cite{Pogorelov1988}. The basic idea is that for small bending stiffness, the shape of the axisymmetric dimple will be close to an \emph{isometric deformation} of the sphere: a shape where a spherical cap is mirror inverted (see fig.\ \ref{fig:midsurface_pog}, grey lines). For vanishing bending stiffness, $E_B=0$, this shape has vanishing elastic energy and, thus, represents the stable equilibrium shape, since it is free of stretching. It only involves bending to invert the curvature of the cap. However, the bending strain at the edge of the inverted cap is infinitely large, which gives rise to infinitely large bending energy for $E_B > 0$. Thus, switching from $E_B=0$ to $E_B>0$, the sharp edge of the dimple has to be smoothed out.

In order to describe the deformation from the isometric shape to the final smooth shape, we follow the ideas of Pogorelov and introduce displacements $u(s_0)$ and $v(s_0)$ in $r$- and $z$-direction, respectively, see fig.\ \ref{fig:midsurface_pog}. Assuming that $u$ and $v$ are small (and some further simplifications), we use linear shell theory to calculate the bending and stretching energies in the final shape. This technique is quite remarkable because it enables us to describe large deformations with linear shell theory by choosing not the undeformed shape as reference state, but the isometric buckled shape. In the following, we will summarise this procedure and results from \cite{Pogorelov1988} regarding the axisymmetric primary buckling because the results will provide the basis for a quantitative theory of the secondary buckling transition.

First, there are geometric relations for isometric deformations of spheres, which are obtained by mirror reflection of a spherical cap, see fig.\ \ref{fig:midsurface_pog} a). The resulting dimple is characterised by its opening angle $\alpha$, from which we can calculate the dimple radius $r_D$, depth $h$ and volume difference $\Delta V$. For later use, we already introduce a first order approximation in $\alpha$, since we will assume that the dimple is small compared to $R_0$. The exact and approximated relations then read
\begin{align}
 r_D &= R_0 \sin \alpha \approx \alpha R_0 \label{eq:r_D} \\
 h &= R_0 - R_0 \cos \alpha \approx \alpha^2 R_0/2 \label{eq:h} \\
 \Delta V &= 2 h^2 \pi (3R_0-h)/3 \approx \pi \alpha^4 R_0^3/2. 
  \label{eq:DeltaV}
\end{align}

In order to evaluate the elastic energies in the final shape, we split it into different regions which are investigated separately. We define the inner neighbourhood $G_i$ and outer neighbourhood $G_o$ of the dimple edge (see fig.\ \ref{fig:midsurface_pog}) as the regions in which the displacements $(u, v)$ are significant. This is only the case in a narrow strip to both sides of the dimple edge, see (\ref{eq:reskal_s0}) below. Outside these regions, the displacements are negligible; these regions are referred to as $G_2$ (for the rest of the dimple) and $G_1$ (for the rest of the undeformed part). 

The elastic energies that we have to evaluate consist of the bending needed to invert the curvature in $G_2$, and the bending and stretching due to the displacements in $G_i$ and $G_o$. The uniform compressive strains in $G_1$ and $G_2$, which result from the negative inner pressure and which are already present in the spherical (pre-buckled) shape, are neglected.
 
The bending energy in $G_2$ is readily written down. From the spherical shape to the isometric deformed shape, the curvatures change from $1/R_0$ to $-1/R_0$, thus giving bending strains $K_s = K_\phi = -2/R_0$, where the stretches in the definition (\ref{eq:bending_strain}) are neglected. The energy density (\ref{eq:w_S}) is integrated over the area of the inverted spherical cap. We neglect that the area of the dimple should be reduced by the area of $G_i$ since we assume that $G_i$ is much smaller than $G_2$, and hence the area of $G_2$ is $A(G_2) = 2 \pi h R_0 \approx \pi \alpha^2 R_0^2$. Thus, the bending energy in $G_2$ reads
\begin{align}
 U_B(G_2) &= \frac{1}{2} E_B \int_{G_2} \diff A 
\left\{ K_s^2 + 2\nu K_s K_\phi + K_\phi^2 \right\} \nonumber \\
 &= 4 \pi \alpha^2 E_B (1+\nu).
\end{align} 

In the region $G_o$, the deformation energy is governed by the displacements $u(s_0)$ and $v(s_0)$ (where $s_0$ is the undeformed arc length). From ``graphic considerations'' \cite{Pogorelov1988} Pogorelov concludes that a meridian will not stretch or compress very much: $e_s \approx 0$. Using  our nonlinear shape equations, we checked that this assumption holds at the secondary buckling transition if the Poisson ratio is not too large; in fact, the magnitudes of $e_s$ and $e_\phi$ (in a root mean square sense) are found to behave roughly as $e_s/e_\phi \sim \nu$ at the secondary buckling transition. At the first buckling transition, we find that $e_s/e_\phi$ depends on $\tilde E_B$, and that our assumption holds for small reduced bending rigidities. The stretching of circumferential fibres (cf.\ eq.\ (\ref{eq:lambda_def})) results in the strain $e_\phi = u(s_0)/r_0(s_0) \approx u(s_0)/r_D$, where the approximation holds when $G_o$ is sufficiently narrow. The integration of the stretching energy is performed over the area element $\diff A \approx 2 \pi r_D \diff s_0$ with $s_0 \in [0, \epsilon]$. Here the arc length coordinate was centred around the dimple edge and runs up to a point $\epsilon$ where the displacements have decayed sufficiently to be neglected. The stretching energy of the outer neighbourhood is thus given by
\begin{align}
 U_S(G_o) &= \frac{1}{2} \frac{EH_0}{1-\nu^2} \int_{G_o} 
       e_\phi^2 \, \diff A \nonumber \\
 &= \frac{\pi EH_0}{\big( 1 - \nu^2 \big) r_D} \int_0^\epsilon  
   u(s_0)^2 \, \diff s_0
\end{align}
Pogorelov approximates the bending strains as $K_s = v''(s_0)$ and $K_\phi = v'(s_0) / \alpha R_0$. In the integral of the bending energy, a term $v'(s_0) v''(s_0)$ occurs, which can be readily integrated to $v'(s_0)^2/2$, and we need to specify boundary conditions for the displacement $v$ to proceed. We require the dimple to have a horizontal tangent at the dimple edge $s_0 = 0$, hence $-v'(0) = \alpha$. At the other end $s_0 = \epsilon$, the displacement shall have decayed and we enforce $v(\epsilon) = v'(\epsilon) = 0$. The resulting expression for the bending energy is 
\begin{equation}
 U_B(G_o) = \pi E_B r_D \int_0^\epsilon  v''(s_0)^2 \, \diff s_0  
            - \pi \alpha E_B \nu r_D / R_0.
\end{equation} 
In total, the elastic energy of the outer neighbourhood $G_o$ is therefore given by
\begin{align}
 U(G_o) = &\int_0^\epsilon \diff s_0 \left\{ \pi E_B r_D v''(s_0)^2 + 
    \frac{\pi EH_0}{\big( 1 - \nu^2 \big) r_D} u(s_0)^2 \right\} \nonumber \\
 &- \pi \alpha E_B \nu r_D / R_0.
 \label{eq:U(G_o)}
\end{align}

Analogously, the elastic energy of the inner neighbourhood $G_i$ can be calculated. The stretching energy is the same as for the outer neighbourhood. But note that the sign of $e_\phi$ is negative in this case because we have hoop compression in the inner neighbourhood and hoop stretching in the outer neighbourhood (we already anticipate that the result for $u(s_0)$ will be positive in the outer neighbourhood, see fig.\ \ref{fig:midsurface_pog}). The bending strains have to be modified because we have to take into account that the curvature of the inverted cap is already inverted, and we get $K_s = v''(s_0) - 2/R_0$ and $K_\phi = -v'(s_0)/\alpha R_0 - 2/R_0$. The resulting elastic energy of the inner neighbourhood is given by
\begin{align}
 U(G_i) = &\int_{-\epsilon}^0 \diff s_0 \left\{ \pi E_B r_D v''(s_0)^2 
   + \frac{\pi EH_0}{\big( 1 - \nu^2 \big) r_D} u(s_0)^2 \right\} \nonumber \\
 &- 4 \pi \alpha E_B (1+\nu) r_D / R_0 + \pi \alpha E_B \nu r_D / R_0.
 \label{eq:U(G_i)}
\end{align}
The integrand coincides with the result (\ref{eq:U(G_o)}) for the outer neighbourhood; only the constant terms differ.

To find the functions $u(s_0)$ and $v(s_0)$ which represent the final shape, we have to minimise the functional of total elastic energy with respect to $u$ and $v$. During the variation $u \rightarrow u+\delta u$ and $v\rightarrow v+\delta v$ we keep the parameters $\alpha$, $h$, $r_D$ and $\Delta V$ of the isometric shape constant. Since the volume change due to $u$ and $v$ can be neglected (compared to the large $\Delta V$ due to the isometric deformation), this corresponds to a variation at constant capsule volume, and we do not need to introduce a Lagrange multiplier controlling the volume. As the integrals in the elastic energies of the inner and outer neighbourhood are identical, we expect a symmetric shape of the dimple, with an odd function $u(s_0)$ and an even function $v(s_0)$. It is thus sufficient to determine the solution on the interval $s_0 \in [0, \epsilon]$ by minimising (\ref{eq:U(G_o)}).

During the minimisation we have to impose a constraint on $u$ and $v$, because the energy functional was set up under the assumption of vanishing (or negligible) meridional strain $e_s = 0$. The final solution must satisfy this condition, which can be written as 
\begin{equation}
 u'(s_0) + \alpha v'(s_0) + \frac{1}{2} v'(s_0)^2 = 0.
 \label{eq:constraint}
\end{equation} 
Furthermore, the variation has to respect the boundary conditions, which are evident from geometrical considerations: $u(0)=0$ so that the capsule is not ripped apart at the dimple edge, $v'(0) = -\alpha$ for a horizontal tangent at the dimple edge, and $u(\epsilon) = v(\epsilon) = v'(\epsilon) = 0$ because the displacements must have decayed at $s_0 = \epsilon$.

The number of parameters in the problem can be greatly reduced with a suitable nondimensionalisation by introducing characteristic length and energy scales. Inspection of the integrand in (\ref{eq:U(G_o)}) and the constraint (\ref{eq:constraint}) shows  that the substitutions
\begin{align}
 s_0&= \xi \bar s_0, \quad
   \xi \equiv \left[ \tilde E_B \big( 1 - \nu^2 \big) \right]^{1/4} \left(\frac{R_0 r_D}{\alpha}\right)^{1/2} \label{eq:reskal_s0}\\
 u  &= \xi \alpha^2 \bar u, & & 
    \label{eq:reskal_u}\\
 \diff v/\diff s_0 &= \alpha \bar w, & &
   \label{eq:reskal_v}
\end{align}
with a typical arc length scale $\xi$ prove useful. For small $\tilde E_B \ll 1$, the length scale $\xi \ll R_0$ is also small, which proves that the regions $G_i$ and $G_o$ are indeed narrow strips. The substitutions lead to a dimensionless form of the energy (\ref{eq:U(G_o)}),
\begin{align}
 U(G_o) &= U_\xi \int_0^{\bar\epsilon} \diff \bar s_0 
 \left\{ \bar w'(\bar s_0)^2 + \bar u(\bar s_0)^2 \right\} + \text{const,}
  \label{eq:U(G_o)2}  \\
 U_\xi &\equiv \pi \alpha^{5/2} r_D^{1/2} R_0^{3/2} 
 \frac{EH_0}{\big( 1 - \nu^2 \big)^{1/4}}  \tilde E_B^{3/4},
\label{eq:Uxi}
\end{align} 
with an energy scale $U_\xi$. Using the geometric relations (\ref{eq:r_D}) - (\ref{eq:DeltaV}), the scaling parameters $\xi$, $U_\xi$, and $\alpha$ can be expressed as functions of the elastic moduli, the reduced volume difference $\Delta V/V_0$, and the capsule radius $R_0$, 
\begin{align}
 \alpha &\approx \left( \frac{8}{3} \frac{\Delta V}{V_0} \right)^{1/4}  
  \label{eq:alpha_linear}\\
 \xi &\approx \left[ \tilde E_B \big( 1 - \nu^2 \big) \right]^{1/4} R_0  
 \label{eq:xi_linear}\\
 U_\xi &\approx \pi \left( \frac{8}{3} \right)^{3/4} 
  \frac{EH_0}{\big( 1 - \nu^2 \big)^{1/4}} 
   \left( \tilde E_B \frac{\Delta V}{V_0} \right)^{3/4} R_0^2.  
 \label{eq:Uxi_linear}
\end{align}
where we used the first-order approximations in $\alpha$. 

Both the  arc length scale $\xi$ and the energy scale   $U_\xi$ emerge from the competition of stretching and bending energies in (\ref{eq:U(G_o)}) (under the  constraint (\ref{eq:constraint})): $\xi$ gives the typical arc length size of the neighbourhoods $G_i$ and $G_o$ and $U_\xi$ gives the typical energy of the buckled configuration. The final result for the Pogorelov buckling energy, which is obtained after minimisation of the total energy  with respect to $u$ and $v$ will differ from $U_\xi$ only by a numerical prefactor, see eq.\ (\ref{eq:Upog}) below.

For the minimisation of the total energy, the integral term in eq.\ (\ref{eq:U(G_o)2}) has to be minimised. Note that the limit $\bar\epsilon$ of the integral has been rescaled, too, according to (\ref{eq:reskal_s0}). Following Pogorelov, we consider the case of small $\tilde E_B$, where  $\bar\epsilon \rightarrow \infty$ because $\xi \rightarrow 0$ according to (\ref{eq:xi_linear}). Thus our task for the calculus of variations is to minimise the functional
\begin{equation}
 J[\bar u, \bar w] = \int_0^\infty \diff \bar s_0 
   \left\{ \bar w'^2 + \bar u^2 \right\}
 \label{eq:J_def}
\end{equation}
subjected to the constraint (\ref{eq:constraint}) which reads 
\begin{equation}
 \bar u' + \bar w + \frac{1}{2} \bar w^2 = 0
 \label{eq:constraint_nd}
\end{equation} 
in rescaled variables and with boundary conditions
\begin{align}
 \bar u(0) = 0,  \quad 
 \bar w(0) = -1, \quad 
 \bar u(\infty) = 0, \quad 
 \bar w(\infty) &= 0.
 \label{eq:BC_nd}
\end{align}

Pogorelov solved the constrained variational problem analytically. His results for the minimising functions $\bar u(\bar s_0)$ and $\bar w(\bar s_0)$ are presented in appendix \ref{app:sol_variation}. These functions are defined piecewise, due to some simplifications, on two intervals $\bar s_0 \in [0,\sigma)$ and $\bar s_0 \in [\sigma,\infty)$, where the optimal choice for $\sigma$ is $\sigma_\text{min} = 1.24667$. The minimal value of the functional is found to be $J_\text{min} = 1.15092$.

Now we can switch back from the nondimensionalised quantities to physical quantities in order to analyse the features of Pogorelov's model, plot solutions and compare them to our results from the nonlinear shape equations. The rescaling of the functions $u$ and $v$ describing the capsule shape is obviously given by (\ref{eq:reskal_s0}) - (\ref{eq:reskal_u}).  We also have to take into account that the origin of $s_0$ was shifted to the dimple edge. In the coordinate system of the nonlinear shape equations, the origin of $s_0$ starts at the south pole of the capsule, and the dimple edge (of the isometric deformation) is located at $s_D = \alpha R_0$. When comparing these solutions, the functions from the Pogorelov model must be shifted.

The displacements $u(s_0)$ and $v(s_0)$ can be used to plot the deformed capsule shape and to calculate further properties, like curvatures and tensions. With the  strain definitions $e_s = 0$ and $e_\phi = u/r_D$, we obtain tensions from the linearised Hookean law, see
(\ref{eq:stress-strain_2D}),
\begin{equation}
 \tau_\phi = \frac{EH_0}{1-\nu^2}  \frac{u(s_0)}{r_D}
 \quad \text{and} \quad 
 \tau_s = \nu \, \tau_\phi.
 \label{eq:pogorelov_tensions}
\end{equation}
The definitions of the bending strains imply curvatures
\begin{align}
 \kappa_s &= \begin{cases}
             -1/R_0 + v''(s_0), & s_0 < s_D \\
             1/R_0 + v''(s_0), & s_0 \geq s_D
            \end{cases} \\
 \text{and} \quad
 \kappa_\phi &= \begin{cases}
             -1/R_0 + v'(s_0)/\alpha R_0, & s_0 < s_D \\
             1/R_0 + v'(s_0)/\alpha R_0, & s_0 \geq s_D
            \end{cases} .
 \label{eq:pogorelov_curvatures}
\end{align} 

The total elastic energy is obtained by adding $U(G_2) + U(G_o) + U(G_i)$. We see that the constant terms cancel each other, and only the integral terms of $U(G_i)$ and $U(G_o)$ survive. Each integral term gives $J_{\rm min} U_\xi$. Thus, our final result for the elastic energy of an axisymmetric buckled capsule with a given volume difference is
\begin{align}
 U_\text{Pog} &= 2 J_{\rm min} U_\xi \label{eq:Upog} \\
  &= 2 \pi J_{\rm min}
   \left( \frac{8}{3} \right)^{3/4} 
   \frac{EH_0}{\big( 1 - \nu^2 \big)^{1/4}} 
     \left( \tilde E_B \frac{\Delta V}{V_0} \right)^{3/4} R_0^2. \nonumber 
\end{align}

\subsection{Comparison of the two models}

\begin{figure}[t]
  \centerline{\includegraphics[width=80mm]{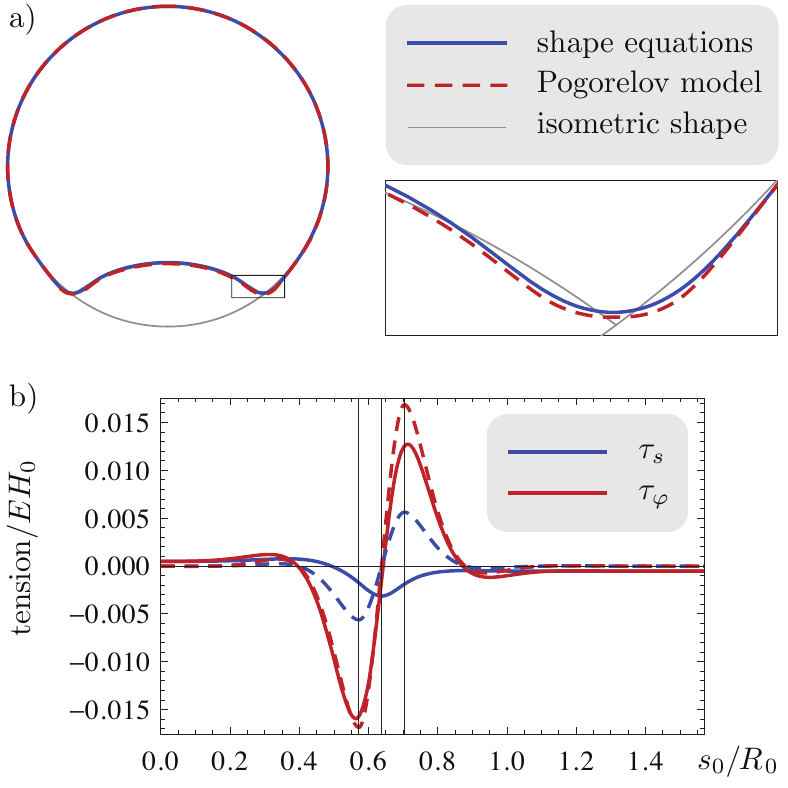}}
  \caption{(Colour online) Comparison of a) shape and b) tension between a solution of the shape equations (continuous lines) and Pogorelov model (dashed lines). The grey line shows the isometric shape. The solutions were calculated for $\Delta V/V_0 = 5\%$, $\tilde E_B = 10^{-5}$, $\nu = 1/3$. The vertical lines indicate the position $s_D$ of the dimple edge and $s_D \pm \sigma_{\rm min} \xi$.}
\label{fig:pogo_comp}
\end{figure}

Figure \ref{fig:pogo_comp} shows a plot of the capsule shape and tension distribution in a solution of the shape equations and a solution of the analytic Pogorelov model. The shape is very well captured by the analytic model. Only in the close-up, deviations from the solution of the shape equations can be recognised. The behaviour of the hoop tension $\tau_\phi(s_0)$ also agrees well in both models. We see a characteristic region of strong compression, which is located in the inner neighbourhood of the dimple edge, in which the models even agree quantitatively. In this region, the compressive hoop tension has a \emph{parabolic} profile to a good approximation. In the outer neighbourhood of the dimple edge, a corresponding region of strong hoop stretch is present. The meridional tension $\tau_s$ is quite badly captured by the Pogorelov model, which is a consequence of the strong simplification of vanishing meridional strain $e_s = 0$.

A simple explanation why hoop compression arises in the inner neighbourhood and hoop tension in the outer neighbourhood of the dimple edge is evident from the close-up in fig.\ \ref{fig:pogo_comp} a): after smoothing of the dimple edge, the inner neighbourhood is displaced horizontally to the left, and the outer neighbourhood to the right. Thus, the circumferential fibres of the inner neighbourhood are compressed and those of the outer neighbourhood are stretched.

\begin{figure}[t]
  \centerline{\includegraphics[width=80mm]{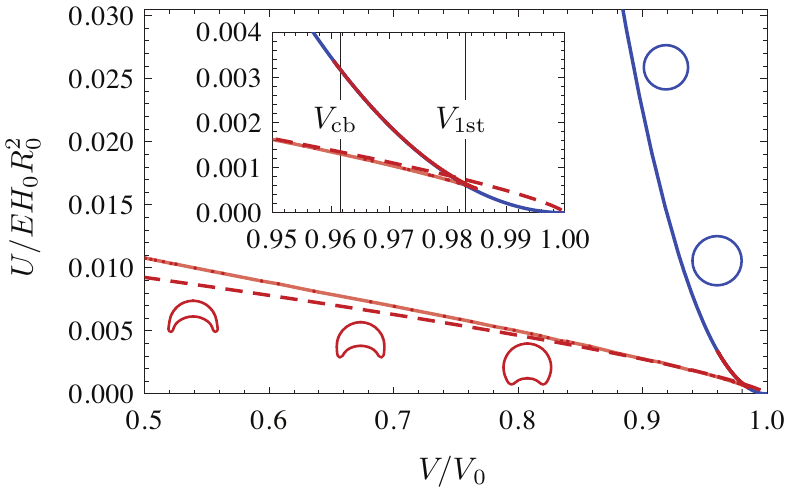}}
  \caption{(Colour online) Elastic energies of the axisymmetric buckled shapes for the shape equations (red points with light red interpolation) and Pogorelov model (dashed line), and for the spherical shapes (blue line). The elastic moduli are $\tilde E_B = 10^{-4}$ and $\nu = 1/3$. In the inset, the critical volume of classical buckling $V_\text{cb}$ according to (\ref{eq:Vcb}) and of the first buckling $V_\text{1st}$ are marked.}
\label{fig:bifdiag_U_V}
\end{figure}

Now, let us turn to the comparison of the elastic energies of the deformed shapes, from which the bifurcation behaviour can be deduced (see \cite{Knoche2011} for bifurcation diagrams of spherical shells). First of all, there is the trivial (spherical) solution branch, which can be handled analytically. If the deformed capsule is a sphere with radius $R$, the strains are uniform and homogeneous, $e_s = e_\phi = R/R_0 - 1 = (V/V_0)^{1/3}-1$. The bending strains vanish. The elastic energy density (\ref{eq:w_S}) must be integrated over the undeformed surface, which has an area $4\pi R_0^2$. Hence, the elastic energy for a spherical shape is given by
\begin{equation}
 U_\text{sph} = 4 \pi \frac{EH_0}{1-\nu} R_0^2 
  \left[ \left( \frac{V}{V_0} \right)^{1/3} - 1 \right]^2.
 \label{eq:Usph}
\end{equation}
This function is plotted in fig.\ \ref{fig:bifdiag_U_V} and increases rapidly with decreasing capsule volume $V$.

The energy of a buckled solution of the shape equations can be calculated by numerical integration of the surface energy density $w_S$ of eq.\ (\ref{eq:w_S}). The data points for numerical solutions are shown in fig.\ \ref{fig:bifdiag_U_V}. At a critical volume $V_\text{cb}$, the classical buckling volume, this solution branch starts to separate from the spherical branch (see inset in fig.\ \ref{fig:bifdiag_U_V}). At first, it runs to the right and lies at slightly higher energies than the spherical solution branch (see the inset). These shapes correspond to unstable energy maxima. But after a return, the buckled branch crosses the spherical branch at a volume $V_\text{1st}$. From there on, the axisymmetric buckled configuration is energetically favourable to the spherical shape, representing the global energy minimum. The spherical shape is still metastable between $V_\text{1st}$ and $V_\text{cb}$ and represents a local energy minimum. Koiter's stability analysis \cite{Koiter1969} suggests that the buckling transition of real (imperfect) shells occurs somewhere in this region, depending on the severity of the imperfections. For $V<V_\text{cb}$, however, the spherical branch is unstable. This bifurcation scenario with metastable spherical and buckled branches below and above $V_{\text{1st}}$, respectively, is typical for a \emph{discontinuous shape transition}. This summarises briefly the discussion of the buckling behaviour presented in ref.\ \cite{Knoche2011}.

In \cite{Knoche2011}, we also showed that our findings for the classical buckling volume coincide with the literature results for the classical buckling pressure $p_\text{cb} = -4 \sqrt{EH_0 E_B}/R_0^2$ (see, for example, \cite{Timoshenko1961,Landau1986,Ventsel2001}). To show that, we need to convert the pressure into a volume, by using the pressure-volume relation of the spherical solution branch. This can be derived from the elastic energy (\ref{eq:Usph}) via
\begin{equation}
 p = \frac{\partial U_\text{sph}}{\partial V} =
   2 \frac{EH_0}{1-\nu} \frac{1}{R_0}
  \left[ \left( \frac{V}{V_0} \right)^{1/3} - 1 \right]
    \left( \frac{V_0}{V} \right)^{2/3}.
\end{equation}
Inverting this relation between $p$ and $V$ and inserting $p=p_\text{cb}$ results in the classical buckling volume
\begin{equation}
 \frac{V_\text{cb}}{V_0} = 
 \left( \frac{1}{2} + \sqrt{\frac{1}{4}+2(1-\nu) \sqrt{\tilde E_B}} \right)^{-3}
 \label{eq:Vcb}
\end{equation} 
or equivalently, for small $\tilde E_B$ \cite{Quilliet2012},
\begin{equation}
 \frac{\Delta V_\text{cb}}{V_0} \approx 6 (1-\nu) \tilde E_B^{1/2}.
 \label{eq:DeltaVcb}
\end{equation} 
This volume coincides very well with the point where the branch of axisymmetric buckled shapes separates from the spherical solution branch (see inset in fig.\ \ref{fig:bifdiag_U_V}).

The elastic energy $U_\text{pog}$ derived in the Pogorelov model, eq.\ (\ref{eq:Upog}), is also incorporated in the bifurcation diagram fig.\ \ref{fig:bifdiag_U_V}. For volumes smaller than $V_\text{1st}$, it agrees well with the data points from the shape equations. Deviations start to develop for large deformations ($V < 0.8 V_0$), which is due to the simplification that the dimple was assumed to be small in the Pogorelov model. For shapes with too small dimples, $V > V_\text{1st}$, the model is also inaccurate, for two reasons. Firstly, it was assumed that the neighbourhood of the dimple which is deformed, $G_i$ and $G_o$, is narrow. For too small dimples, this condition is not satisfied since the size of the inner neighbourhood becomes as large as the dimple itself. Secondly, for shapes with small dimples, the pre-buckling deformation (that is, the uniform contraction due to the negative inner pressure) cannot be neglected as it was done in the Pogorelov model. This is also the reason why the tensions $\tau_s$ and $\tau_\phi$ do not reach the correct limits far away from the dimple edge (see fig.\ \ref{fig:pogo_comp} b).

However, the inset in fig.\ \ref{fig:bifdiag_U_V} suggests that the Pogorelov model can be successfully used to calculate the first buckling volume $V_\text{1st}$ by setting $U_\text{Pog} = U_\text{sph}$. Doing so using (\ref{eq:Upog}) and (\ref{eq:Usph}), we obtain for small $\Delta V/V_0$ the critical volume difference for the first buckling,
\begin{equation}
 \left.\frac{\Delta V_\text{1st}}{V_0}\right|_\text{Pog} 
 = 6 J_{\rm min}^{4/5} 
  \frac{(1-\nu)^{4/5}}{\big( 1 - \nu^2 \big)^{1/5}}  \tilde E_B^{3/5}.
 \label{eq:DeltaV1st}
\end{equation} 

Equations (\ref{eq:DeltaVcb}) and (\ref{eq:DeltaV1st}) define two lines in a phase or stability diagram, see fig.\ \ref{fig:phase_diag}, which is set up in the $\Delta V$-$\tilde E_B$-plane. In the phase diagram, the line of the first buckling transition is represented by a continuous line according to the Pogorelov model (\ref{eq:DeltaV1st}) and data points which were derived from the shape equations by requiring equal energies of the spherical and buckled shape. Both approaches are in good agreement. The data points can be fitted with a power law (see fig.\ \ref{fig:phase_diag}),
\begin{equation}
 \left. \frac{\Delta V_\text{1st}}{V_0} \right|_\text{shape eqs.} 
    = (4.78\pm0.03)\, \tilde E_B^{0.6127 \pm 0.0006}
\end{equation} 
This must be compared to (\ref{eq:DeltaV1st}) (evaluated at $\nu = 1/3$ as used in fig.\ \ref{fig:phase_diag}), which is $4.97 \, \tilde E_B^{0.6}$ and thus very close. The classical buckling transition is also represented by a continuous line, according to (\ref{eq:Vcb}), and data points from the shape equations in the phase diagram fig.\ \ref{fig:phase_diag}. In the space between the lines of first and classical buckling, spherical shapes and axisymmetric buckled shapes can both exist, since the spherical shapes are metastable and the buckled shapes stable.

With that we close our investigation of the axisymmetric shapes. In conclusion, we have shown that the hoop tension $\tau_\phi(s_0)$ has a negative peak in the inner neighbourhood of the dimple edge. This is the region where wrinkles will develop when the compressive tension exceeds a critical value. Furthermore, we have established two critical volumes for the transition from the spherical shape to the axisymmetric buckled shape. These are the first buckling volume $V_\text{1st}$, at which the buckled shape becomes energetically favourable to the spherical shape, and the classical buckling volume $V_\text{cb}$ where the spherical shape loses its (meta)stability.

\section{Secondary buckling as 
wrinkling under locally compressive stress}
\label{sec:secbuck_wrinkling}

\subsection{Simplification of geometry and stress state}

One main result of the previous section is that close to the dimple edge, a region of compressive hoop tension $\tau_\phi$ with a parabolic profile occurs, see fig.\ \ref{fig:pogo_comp} b). This motivates our analysis of wrinkling or buckling of an elastic plate under a locally compressive parabolic stress in this section: We expect wrinkles to occur in the region of compressive hoop stress because the formation of wrinkles can release the compressive stress, which is energetically favourable \cite{Cerda2003,Wong2006}. Within this relevant region, the capsule is shallow, and we can approximate it as a \emph{shallow shell} or \emph{curved plate}.

\begin{figure}
  \centerline{\includegraphics[width=80mm]{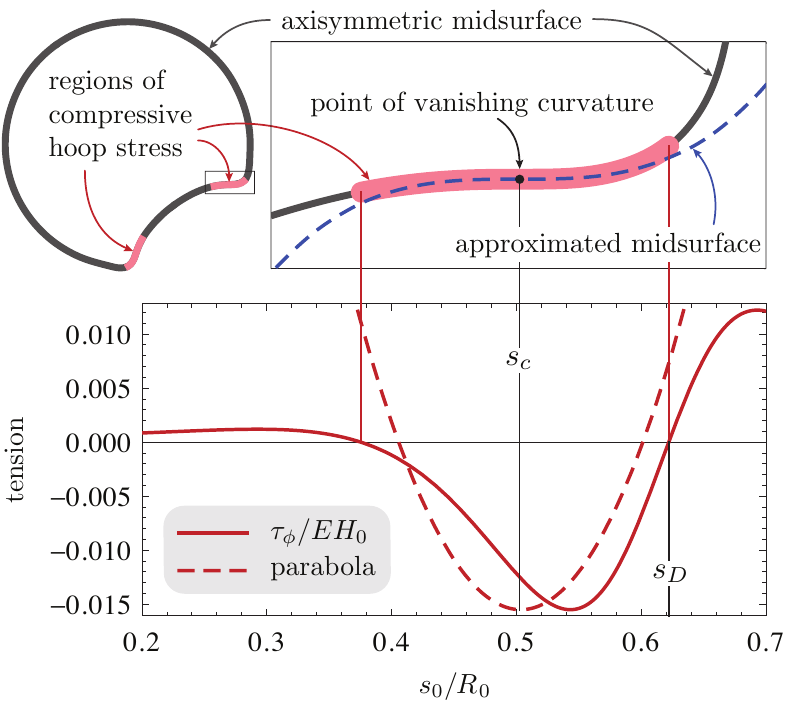}}
  \caption{(Colour online) Shape and stress distribution for a numerical solution of the shape equations at $\tilde E_B = 10^{-5}$, $\nu=1/3$ and $\Delta V / V_0 = 5\%$. The hoop tension $\tau_\phi$ is compressive (negative) in a narrow strip along the inner side of the dimple edge, marked light red, where we can expect wrinkles to occur. The dashed lines show the simplifications made for the stability analysis, namely a parabolic stress profile and a cubic parabola for the approximated midsurface.}
\label{fig:compressive_tension}
\end{figure}

The specific plate geometry and the state of stress we impose are shown in fig.\ \ref{fig:compressive_tension}. The key features of the stress distribution and midsurface geometry in the wrinkling region can be reduced to the following simple functions: The stress distribution $\tau_\phi(s_0)$ can be approximated by a parabola (dashed red line), and the section through the midsurface by a cubic parabola (dashed blue line). Note that these approximations only have to hold in the wrinkled region, i.e.\ the region of compressive $\tau_\phi$, or, at the end of our analysis, the region with a large wrinkle amplitude. We need three parameters to describe the state of stress and the plate geometry in the following: the parameters $\tau_0$ and $a_p$ characterise the depth and width of the parabolic stress profile, and the parameter $a_c$ the curvature of the section through the midsurface.

The cubic parabola describing the midsurface is fitted to the point where the exact midsurface has vanishing curvature $\kappa_s(s_c)=0$ (see fig.\ \ref{fig:compressive_tension}). In the vicinity of this point, the real midsurface shows an approximately linear increase in curvature, $\kappa_s(s_0) \approx a_c (s_0 - s_c)$. To obtain a cubic parabola with the same slope of curvature, we choose a height profile
\begin{equation}
 z(x, y) = \frac{1}{6} a_c y^3
 \label{eq:z(x,y)}
\end{equation} 
to describe the approximated midsurface. Here we have oriented the $y$-coordinate along the original $s_0$-coordinate and centred it at the point of vanishing curvature. By writing down (\ref{eq:z(x,y)}), we have neglected the radius of curvature $1/\kappa_\phi$ of the axisymmetric solution, because we assume that the wrinkles have shorter wavelength. Therefore, we consider a  curved rectangular plate with  the $\phi$-coordinate corresponding to the uncurved $x$-coordinate. Since the relevant portion of the shell is shallow, differences in the metric for the description with $s_0$ or $y$ as a coordinate can also be neglected \cite{Ventsel2001}. So we can calculate $a_c$ by simple differentiation of $\kappa_s(s_0)$ with respect to $s_0$ instead of $y$, which can be done numerically for a given axisymmetric solution,
\begin{equation}
 a_c = \left.\frac{\diff \kappa_s}{\diff s_0}\right|_{s_c}.
 \label{eq:alpha_c}
\end{equation}

The parabola to approximate the hoop tension is chosen to have the same minimum value $-\tau_0$ and the same integral $\int \tau_\phi \, \diff s_0$ over the compressive part as the exact numerical function $\tau_\phi(s_0)$. Let $F=\int_{s_1}^{s_2} \tau_\phi(s_0) \, \diff s_0$ denote the exact numerical integral, which has the physical interpretation of the net force in the compressive region $s_0 \in [s_1, s_2]$. We can evaluate it by numerical integration for a given solution. A parabola of the form
\begin{equation}
 \tau_{x} (x, y) =-\tau_0\left( 1-a_p y^2 \right) 
 \label{eq:tau_x(x,y)}
\end{equation} 
has the roots $y = \pm 1/\sqrt{a_p}$. The integral over the parabola between its roots is $-4\tau_0/3\sqrt{a_p}$ and must equal $F$. Thus we have
\begin{equation}
 a_p = (4\tau_0/3F)^2 \quad \text{with}
   \quad F=\int_{s_1}^{s_2} \tau_\phi(s_0) \, \diff s_0
 \label{eq:alpha_p} 
\end{equation}
to determine the parameter of the parabola from a given axisymmetric shape. Note that the parabola is centred at $y = 0$, corresponding to the point $s_0 = s_c$ of vanishing meridional curvature. This point does not exactly agree with the minimum of the exact hoop tension $\tau_\phi(s_0)$, but is very close (see fig.\ \ref{fig:compressive_tension}).

In the following, we will neglect the meridional tension $\tau_s$, since it is small compared to $\tau_\phi$, see fig.\ \ref{fig:pogo_comp}. Furthermore, there are no shear tensions in the axisymmetric configuration. Hence the stress state to which the plate is subjected reads $\tau_y = 0$, $\tau_{xy} = 0$ and $\tau_x$ according to (\ref{eq:tau_x(x,y)}).

\begin{figure}
  \centerline{\includegraphics[width=80mm]{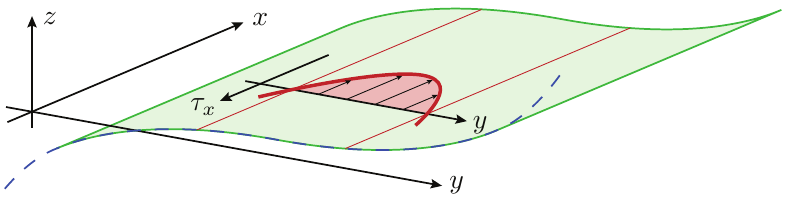}}
  \caption{(Colour online) The simplified stability problem: A plate (curved in $y$-direction) is subjected to a parabolic stress profile $\tau_x$ and will wrinkle in the region of compressive stress.}
\label{fig:plate}
\end{figure}

Figure \ref{fig:plate} summarises our simplified model. We want to investigate the wrinkling (or buckling) of a curved plate in the $(x, y)$ plane with height profile (\ref{eq:z(x,y)}), subjected to a locally compressive stress (\ref{eq:tau_x(x,y)}). The appropriate tool for this task are the \emph{stability equations of shallow shells}. The result of this analysis  will be a wrinkling criterion in the form of a critical tension $\tau_c$ such that for compressive stress levels $\tau_0>\tau_c$ the plate wrinkles. The critical tension $\tau_c$ will depend on the parameters $a_p$ characterizing the width of the parabolic stress profile and  $a_c$ characterizing the curvature of the midsurface section, and the elastic moduli of the shell.

\subsection{Wrinkling criterion for the curved plate}
\label{sec:wrinkling_crit_curved_plate}

Before presenting a more detailed stability analysis, we start with a scaling argument. Here we neglect curvature effects completely and approximate the compressed region of the plate by a rectangular strip of width $\Delta y \sim 1/\sqrt{a_p}$ (identical to the compressive part of the parabola) under a homogeneous compressive stress $\tau_x \sim -\tau_0$. For clamped long edges, the wrinkling wave length is given by the width, $\lambda \sim \Delta y$ \cite{Timoshenko1961}, and the resulting critical Euler buckling stress is $\tau_0 = \tau_c \sim E_B/\lambda^2 \sim E_B a_p$. This result turns out to give the correct parameter dependence in leading order, see eq.\ (\ref{eq:tau_crit}) below.

The stability equations are partial differential equations for the normal displacement $w(x,y)$ and Airy stress function $\sphi(x,y)$ and are given by
\cite{Ventsel2001}
\begin{align}
 \Lap^2 \sphi &= - EH_0 \vec\nabla_\kappa^2 w \label{eq:stability_1}\\
 E_B \Lap^2 w &= \vec\nabla_\kappa^2 \sphi + \tau_x \del_{xx}w 
+ 2 \tau_{xy} \del_{xy} w + \tau_y \del_{yy} w,
\label{eq:stability_2}
\end{align} 
see appendix \ref{app:wrinkling} for a derivation.
Here, $\Lap = \del_{x}^2 + \del_{y}^2$ is the Laplacian and $\vec\nabla_\kappa^2 = \kappa_y \del_{x}^2 + \kappa_x \del_{y}^2$ is the Vlasov operator.
Note that the tensions $\tau_i$ and curvatures $\kappa_i$ occurring in these equations are the tensions and curvatures of the initial state, prior to wrinkling. The Airy stress function, or stress potential, permits the calculation of the additional stresses in the plate, which arise because of the wrinkling, by the relations $\tau_x^{(1)} = \del_y^2 \sphi$, $\tau_{xy}^{(1)} = -\del_x \del_y \sphi$ and $\tau_y^{(1)} = \del_x^2 \sphi$ \cite{Ventsel2001}. The existence of a non-trivial solution of these  stability equations indicates the existence of an unstable deformation mode for  the axisymmetric buckled solution (i.e.\ a negative eigenvalue of the second variation of the elastic energy).

In the present geometry and stress state, some terms vanish in the stability equations. We solve the equations using an ansatz
\begin{equation}
 w(x,y) = W(y) \sin kx, \quad \sphi(x,y) = \Phi(y) \sin kx.
\label{eq:ansatz_stabeq}
\end{equation}
The wrinkle shape function $w(x,y)$ represents a wrinkle pattern consisting of wrinkles extending in $y$-direction with an amplitude function $W(y)$ for each wrinkle,  which are arranged in a periodic pattern in $x$-direction with a wave number $k$. Wrinkle shape $W(y)$ and wave number $k$ are to be determined. Inserting this ansatz as well as the expressions for tensions and curvatures into the stability equations results in two coupled linear ordinary differential equations for the amplitude functions,
\begin{align}
 0 &= \left(\del_y^4 - 2k^2 \del_y^2 + k^4-\frac{k^2}{E_B}\tau_0
 \left( 1-a_p y^2 \right) \right) W +
   \frac{a_c k^2 y}{E_B} \Phi \nonumber \\
 0 &= \left( \del_y^4 - 2 k^2 \del_y^2 + k^4 \right) \Phi 
  - \left(a_c y k^2 EH_0\right) W. 
 \label{eq:DGL}
\end{align}

For a numerical solution, it is necessary to nondimensionalise the equations, which gives useful information on the relevant parameters. At first, we choose a length unit $1/\sqrt{a_p}$, which is the root of the parabolic stress profile, so that we can expect $W(y)$ to decay on this scale. Substituting $y = \hat y / \sqrt{a_p}$ and $\del_y = \sqrt{a_p} \,  \del_{\hat y}$ in (\ref{eq:DGL}) induces further substitutions for the parameters of the differential equations such that they can finally be written in the form
\begin{align}
 0 &= \left(\del_{\hat y}^4 - 2 \hat k^2 \del_{\hat y}^2 + \hat k^4 
  - \hat k^2 \hat\tau_0 \left( 1-\hat y^2 \right) \right) \hat W 
  + \left( \hat a_c \hat k^2 \hat y \right) \hat \Phi \nonumber \\
 0 &= \left( \del_{\hat y}^4 - 2 \hat k^2 \del_{\hat y}^2 + \hat k^4 \right) 
  \hat \Phi - \left( \hat a_c \hat k^2 \hat y \right) \hat W. 
  \label{eq:DGL_nondim}
\end{align}
The substitutions included here are
\begin{equation}
\begin{aligned}
 \hat y &= \sqrt{a_p}  y, & 
 \hat k &= \frac{k}{\sqrt{a_p}}, & 
 \hat a_c &= \sqrt{\frac{EH_0}{E_B}} \frac{a_c}{a_p^{3/2}}, \\
 \hat \tau_0 &= \frac{\tau_0}{a_p E_B}, &
 \hat \Phi &= \frac{\Phi}{E_B}, &
 \hat W &= \sqrt{\frac{EH_0}{E_B}}  W.
 \label{eq:nondim}
\end{aligned}
\end{equation} 

A shooting method \cite{Stoer2010,numrec} can be applied to solve the differential equations numerically, when boundary conditions are provided. Because of the symmetry of the problem, we expect $\hat W(\hat y)$ to be an even function. Then, from (\ref{eq:DGL_nondim}), it follows that $\hat \Phi(\hat y)$ is an odd function. Thus, it is sufficient to solve the differential equations on an interval $0 \leq \hat y \leq \hat y_\text{max}$. The boundary conditions at the left end of this interval, the start-point of integration, follow from the symmetry conditions, $\hat\Phi(0) = \hat\Phi''(0) = 0$, $\hat W'(0) = \hat W'''(0) = 0$ and $\hat W(0) = 1$. The latter choice is arbitrary, since the differential equations are homogeneous. We imagine the plate to be infinitely large, so that the wrinkles are confined by the local nature  of the compression rather than  plate edges. For $\hat y \rightarrow \infty$, the wrinkle amplitude $\hat W$ has to approach $0$, as well as the slope $\hat\Phi'$ of the stress potential because the additional tension derived from $\Phi$ shall approach $0$. In practice, we integrate up to a sufficiently large $\hat y_\text{max}$ and impose the boundary conditions $\hat W(\hat y_\text{max}) = \hat W'(\hat y_\text{max}) = 0$ and $\hat \Phi'(\hat y_\text{max}) = \hat \Phi''(\hat y_\text{max}) = 0$. When the shooting method must satisfy four boundary conditions at the endpoint of integration, it needs to vary four independent shooting parameters, which are typically the starting values of integration. But in the present case, due to the homogeneity of the differential equations, there are only three free starting conditions, $\hat W''(0)$, $\hat \Phi'(0)$ and $\hat \Phi'''(0)$; the choice of $W(0)$ is arbitrary and cannot serve as a shooting parameter. Thus, the shooting method must be allowed to vary one of the additional parameters of the differential equations, $\hat k$, $\hat a_c$ or $\hat \tau_0$.

In fact, the differential equations can be interpreted as an \emph{eigenvalue problem}, where one of the three parameters  $\hat k$, $\hat a_c$ or $\hat \tau_0$ plays the role of an  ``eigenvalue'' and must be chosen so that the equations do have a non-trivial solution. In our case, we choose $\hat \tau_0$ as the eigenvalue, because this has the simplest physical interpretation: we increase the stress on the plate until a non-trivial solution in form of wrinkles exists. Then, this value is the critical stress $\hat \tau_0$ which the plate can bear; for larger loads it will wrinkle. Obviously, the critical stress will depend on the other two parameters, $\hat \tau_0 = \hat \tau_0(\hat a_c, \hat k)$. The value of $\hat a_c$ is known when an axisymmetric solution is given, see (\ref{eq:alpha_p}) and (\ref{eq:alpha_c}). The wave number $\hat k$, however, is unknown. We are interested in the wrinkling mode which becomes unstable first, i.e.\ for the smallest possible load. To find this critical mode and the corresponding critical stress, we can minimise $\hat \tau_0$ with respect to $\hat k$,
\begin{equation}
 \hat \tau_c(\hat a_c) = \min_{\hat k} \hat \tau_0(\hat a_c, \hat k),
\end{equation} 
so the non-dimensional critical stress only depends on $\hat a_c$. The wavenumber $\hat k_c$ of the minimum is related to the wavelength of the wrinkles, $\hat \lambda_c = 2\pi/\hat k_c$. Due to boundary conditions in $x$-direction, the wavelength is quantised, and thus the minimisation also has to be performed on a quantised space for $\hat k$. For simplicity, however, we will neglect the quantisation, which corresponds to an infinitely long plate (in $x$-direction) on which all wavelengths are permitted.

\begin{figure}
  \centerline{\includegraphics[width=80mm]{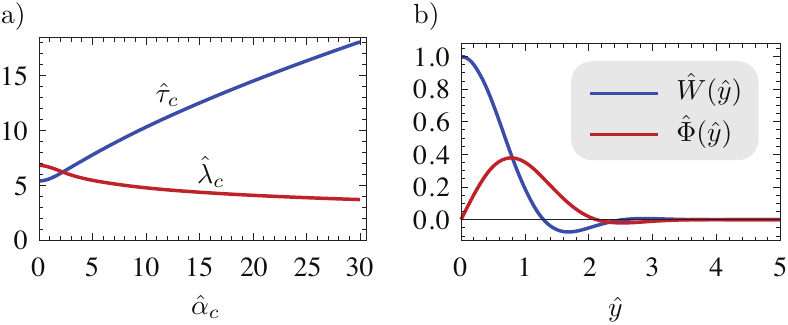}}
  \caption{(Colour online) a) Numerical results for the non-dimensional critical tension $\hat \tau_c = \tau_c/E_B a_p$ and wavelength $\hat\lambda_c = \lambda_c \sqrt{a_p}$ as a function of the curvature parameter $\hat a_c = \sqrt{EH_0/E_B}  a_c/a_p^{3/2}$. b) Numerical solution for the wrinkle amplitude functions for $\hat a_c = 20$.}
\label{fig:krumm_results}
\end{figure}

A numerical analysis along the lines just presented reveals the function $\hat \tau_c (\hat a_c)$, which is shown in fig.\ \ref{fig:krumm_results} a) together with the critical wavelength. The critical tension increases with increasing curvature parameter $\hat a_c$. This reflects the well known fact that bent surfaces (like corrugated cardboard) are harder to bend in the perpendicular direction than flat surfaces. In addition, our numerical procedures return the shape of the wrinkles, that is, the amplitude function $\hat W(\hat y)$. As shown in fig.\ \ref{fig:krumm_results} b), the amplitude indeed decays rapidly outside the compressive region, i.e.\ for $|\hat y| > 1$, as it is required for the approximation to be accurate.

The main result of this section is the formula for the critical compressive stress on which a curved plate will start to wrinkle,
\begin{equation}
 \tau_c = a_p E_B \hat\tau_c(\hat a_c) 
 \quad \text{with} \quad 
 \hat a_c = \sqrt{\frac{EH_0}{E_B}}  \frac{a_c}{a_p^{3/2}}.
 \label{eq:tau_crit}
\end{equation} 
The function $\hat \tau_c(\hat a_c)$ is known numerically (see fig.\ \ref{fig:krumm_results} a). For a given axisymmetric solution of the shape equations (or the Pogorelov model), the parameters $a_c$ and $a_p$ can be calculated according to (\ref{eq:alpha_c}) and (\ref{eq:alpha_p}), respectively, and inserted into (\ref{eq:tau_crit}). This gives the critical tension which the capsule can bear in the axisymmetric state. If the minimum value $\tau_\text{min} = \min_{s_0}\tau_\phi(s_0)$ in the compressive region exceeds the critical stress, i.e.\ if $|\tau_\text{min}| > \tau_c$, the axisymmetric capsule shape is unstable with respect to a wrinkling mode with wavelength
\begin{equation}
 \lambda_c = \frac{1}{\sqrt{a_p}} \, \hat \lambda_c(\hat a_c),
 \label{eq:lambda_crit}
\end{equation} 
where the function $\hat \lambda_c(\hat a_c)$ is also known numerically (see fig.\ \ref{fig:krumm_results} a). This is our \emph{secondary buckling criterion} according to the curved plate model.

Our further analysis also shows that the secondary buckling transition is a \emph{continuous} transition, see appendix \ref{app:wrinkling}.
We checked that the fourth order terms in the elastic energy are positive, and that the energy change upon wrinkling reads $\Delta \hat U \sim -(\hat\tau_0 - \hat\tau_c) \hat W_0^2 + \hat W_0^4$ for a given wrinkle amplitude $\hat W_0\equiv\hat W(0)$ (all numerical prefactors have been omitted). Thus, if the compressive tension exceeds the critical tension, $\tau_0 > \tau_c$, the formation of wrinkles lowers the elastic energy in second order,
which yields a wrinkle amplitude growing like $\hat W_0 \sim \sqrt{\hat \tau_0 - \hat \tau_c}$ when the critical tension is exceeded. At the secondary buckling transition the system thus undergoes a \emph{supercritical pitchfork bifurcation}. The continuity of the secondary buckling is also observed in the numerical simulations \cite{Quilliet2012}. This is in contrast to the primary buckling transition, which is a \emph{discontinuous} transition with metastability above and below the transition as discussed above and with an axisymmetric dimple of the buckled state which always has a finite size.

\subsection{Secondary buckling transition threshold}

We will now apply the secondary buckling criterion to buckled shapes from the
shape equations and from the Pogorelov model in order to quantify the
threshold for secondary buckling.

\begin{figure}
  \centerline{\includegraphics[width=80mm]{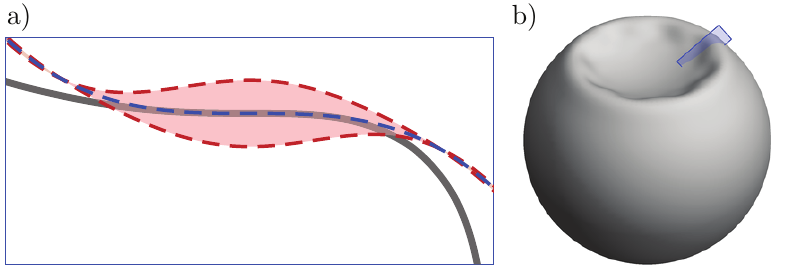}}
  \caption{(Colour online) Wrinkle shape according to the curved plate model for $\tilde E_B = 10^{-5}$ and $\nu = 1/3$ at the critical volume for the secondary buckling, the wrinkle amplitude is arbitrary. a) Section through the midsurface. Axisymmetric midsurface in grey, approximating cubic parabola in dashed blue, and wrinkle amplitude $W$ in dashed red. b) A three-dimensional view, the blue frame indicates the position and orientation of the section shown in a).}
\label{fig:wrinkles_curved_plate}
\end{figure}

 For the shape equations, we exemplarily discuss a capsule with elastic parameters $\tilde E_B = 10^{-5}$ and $\nu = 1/3$ (see fig.\ \ref{fig:wrinkles_curved_plate}). Our numerical analysis shows that the critical volume difference for the secondary buckling is $4.8\%$, for the dimensionless curvature parameter we find $\hat a_c = 20.86$, and the wrinkle wavelength is $\lambda_c = 0.396 R_0$. The wrinkle wavelength can be converted into a wrinkle number: Since the wrinkles are centred at $s_c$ where the radius of the axisymmetric midsurface is $r(s_c) = 0.476 R_0$, the number of wrinkles is given by $n = 2\pi r(s_c) / \lambda_c = 7.55$, so either $n=7$ or $n=8$. These results are illustrated in fig.\ \ref{fig:wrinkles_curved_plate} in a meridional section and in a three-dimensional view. In fig.\ \ref{fig:wrinkles_curved_plate} a), the same section as in fig.\ \ref{fig:compressive_tension} is shown, but additionally the normal displacement, or wrinkle amplitude, is plotted. The results show that the wrinkle amplitude is very small outside the compressive region (compare with fig.\ \ref{fig:compressive_tension}), and that the approximated midsurface is quite accurate in the region of large wrinkle amplitude. This indicates that our approximations are justified, because their inaccuracies lie in regions where the wrinkles do not develop.

We have applied this analysis to a whole range of different bending stiffnesses $\tilde E_B$, and thus generated a new line for the phase diagram fig.\ \ref{fig:phase_diag} which represents the critical volume of the secondary buckling transition (red data points). The data points can be fitted by the power law
\begin{equation}
 \left. \frac{\Delta V_\text{2nd}}{V_0} \right|_\text{shape eqs.} 
     =  (2550 \pm 50) \, \tilde E_B^{0.946 \pm 0.002}.
 \label{eq:DeltaV2ndfit}
\end{equation} 

Now, the secondary buckling criterion will be applied to the Pogorelov model. Because we have an approximate analytic expression for the compressive hoop stress profile within the Pogorelov model, we can obtain an analytic result for the secondary buckling threshold. At first, the parameters $a_p$ and $a_c$ must be determined, so that we can evaluate the critical hoop tension (\ref{eq:tau_crit}). In the Pogorelov model, the hoop tension $\tau_\phi$ is given by (\ref{eq:pogorelov_tensions}). Its minimum value is
\begin{equation}
 \label{eq:tau_0_pogo}
 \tau_0 = \frac{\sigma_{\rm min}}{3} \frac{EH_0}{1-\nu^2} \frac{\xi \alpha}{R_0}, 
\end{equation} 
and the integral between its roots can be calculated as
\begin{equation}
 F = \frac{EH_0}{1-\nu^2} \frac{\xi^2\alpha}{R_0} 
  \left[ \frac{5}{24}\sigma_{\rm min}^2 + \frac{\sigma_{\rm min}}{3} 
    \left( e^{-3\pi/4} + \sqrt{2} \right) \right].
\end{equation} 
Using eq.\  (\ref{eq:alpha_p}), it can be shown that
\begin{equation} \label{eq:a_p_Pog}
  a_p = \frac{\bar a_p}
      {R_0^2 \big[\tilde E_B \big( 1 - \nu^2 \big)\big]^{1/2}}
  \quad \text{with} \quad 
  \bar a_p \approx 0.33955
\end{equation}
where, for convenience, the numerical constants have been reduced to $\bar a_p$, and the linearised scaling parameters (\ref{eq:alpha_linear}) - (\ref{eq:Uxi_linear}) were used.

Calculating the curvature parameter $a_c$ is a bit more difficult and cannot be done explicitly. The problem is that the root of the meridional curvature according to (\ref{eq:pogorelov_curvatures}) cannot be found analytically. It is implicitly determined by the equation
\begin{equation}
 \bar w'(\bar s_c) - \frac{\xi}{\alpha R_0} = 0.
\end{equation} 
Inserting the linearised scaling parameters and introducing the substitution $X = (\Delta V/V_0 \tilde E_B)^{1/4}$, this equation reads
\begin{equation}
 \bar w'(\bar s_c) - \left( \frac{3}{8} \right)^{1/4} 
   \big( 1 - \nu^2 \big)^{1/4}  \frac{1}{X} = 0
\end{equation} 
and implicitly defines the function $\bar s_c(X)$, which can be evaluated numerically, for example by using a Newton method. The curvature parameter $a_c$, see (\ref{eq:alpha_c}), can then also be evaluated numerically for given $X$,
\begin{equation}
 a_c(X) = \frac{\alpha}{\xi^2}  \bar w''(\bar s_c(X)).
\end{equation} 
This leads to the nondimensionalised curvature parameter, see
(\ref{eq:tau_crit}),
\begin{equation}
 \hat a_c(X) = \left( \frac{8}{3} \big(1-\nu^2\big) \right)^{1/4} 
  \frac{X}{\bar a_p^{3/2}}  \, \bar w''(\bar s_c(X)).
\end{equation}

After these steps, the critical tension $\tau_c$ (\ref{eq:tau_crit}) can be evaluated, as a function of $X$. The critical volume difference for the secondary buckling requires $\tau_c = \tau_0$ with $\tau_0$ from (\ref{eq:tau_0_pogo}), which is equivalent to
\begin{equation}
 \hat \tau_c(\hat a_c(X)) = 
  \frac{\sigma_{\rm min}}{3 \bar a_p} 
 \left( \frac{8}{3} \right)^{1/4} \big(1-\nu^2\big)^{-1/4}  X.
\end{equation}
Solving this equation numerically results, for $\nu = 1/3$, in $X=7.8024$. Thus, with the original definition $X = (\Delta V/V_0 \tilde E_B)^{1/4}$, we have derived the secondary buckling volume
\begin{equation}
 \left. \frac{\Delta V_\text{2nd}}{V_0} \right|_\text{Pog}
   = 3706\,  \tilde E_B
 \label{eq:DeltaV2nd_Pogorelov}
\end{equation} 
for the Pogorelov model with the curved plate buckling criterion. In the phase diagram fig.\ \ref{fig:phase_diag}, this line (in red) is close to the second-buckling-line (\ref{eq:DeltaV2ndfit}) of the shape equations, but has a slightly different exponent. The exponent $1$ is exact in the Pogorelov model, but the prefactor in (\ref{eq:DeltaV2nd_Pogorelov}) is still weakly $\nu$-dependent.

From these results, the wrinkle wavelength can also be calculated. At $X=7.8024$, we have (always for $\nu=1/3$) a curvature parameter of $\hat a_c(X) = 15.14$ and a non-dimensional wavelength of $\hat \lambda_c(\hat a_c) = 4.366$ (see fig.\ \ref{fig:krumm_results}). This results in a real wavelength of
\begin{equation}
 \lambda_c = \frac{\hat \lambda_c}{\sqrt{a_p}} 
  = R_0 \big[\tilde E_B  \big(1-\nu^2\big) \big]^{1/4} 
   \frac{ \hat \lambda_c}{\sqrt{\bar a_p} }.
\end{equation} 
For the number of wrinkles which are distributed along the perimeter 
$2 \pi r_D$ we then get\footnote{The result $n \approx 8.6$ corrects an erroneous statement in ref.~\cite{Knoche2014}}
\begin{equation}
 n = \frac{2\pi r_D}{\lambda_c} = 
  2 \pi \left( \frac{8}{3} \right)^{1/4} 
  \frac{X \sqrt{\bar a_p}}{\hat \lambda_c \big( 1-\nu^2 \big)^{1/4}} 
   \approx \Cr{8.6} 
 \label{eq:Pogorelov_wrinkle_number}
\end{equation} 
Thus, in this combination of simplified models (Pogorelov model for the axisymmetric shape and buckling criterion from the curved plate model), the number of wrinkles seems fixed over the whole range of $\tilde E_B$.

\section{Secondary buckling from a stability analysis of the full axisymmetric buckled shape}

To confirm the results for the secondary buckling transition just presented, we also followed another, more rigorous approach. The stability equations of shallow shells, which we already used for the secondary buckling criterion for the curved plate, can also be directly applied to the full axisymmetric shape. In comparison with the curved plate model, this is an improvement because now, also the less prominent features like the meridional tension $\tau_s \neq 0$ and circumferential curvature $\kappa_\phi \neq 0$ are contained in the stability analysis. However, it might be problematic that the dimple is not shallow for large volume differences. In the curved plate model, this problem was avoided because we only looked at a small section of the shell, which was even for large dimples quite shallow.

\subsection{Stability equations of shells of revolution}

The stability equations of shallow shells presented in section \ref{sec:wrinkling_crit_curved_plate} are formulated for Cartesian coordinates, which are not appropriate for our case of an axisymmetric shape (before wrinkling). We did not find a formulation of the stability equations in general coordinates in the shell theory literature. However, the derivation of the Cartesian stability equations (appendix \ref{app:wrinkling}) can be transferred to a suitable coordinate system. The basis is formed by the strain-displacement relations of the nonlinear DMV theory, a simplified shell theory named after Donnel, Mushtari and Vlasov. A formulation of this theory in general coordinates can be found in ref.\ \cite{Niordson1985}. Appendix \ref{app:stab_DMV} shows that the stability equations can be formulated again in the normal displacement $w$ and stress potential $\sphi$ and are formally very similar to the Cartesian case,
\begin{align}
  E_B  \Lap^2 w &= (\kappa_s D_{\phi\phi} + \kappa_\phi D_{ss}) \sphi 
 +  (\tau_s D_{ss} + \tau_\phi D_{\phi\phi}) w \nonumber \\
  \frac{1}{EH_0}  \Lap^2 \sphi &= 
    -(\kappa_s D_{\phi\phi} + \kappa_\phi D_{ss}) w.
  \label{eq:DMV_stability}
\end{align}
Only the derivative operators must be re-defined according to
\begin{equation}
\label{eq:Dss_etc}
\begin{aligned}
 &D_{ss} \equiv \del_{s}^2, \quad 
 D_{\phi\phi} \equiv \frac{\cos \psi}{r} \, \del_{s} 
   + \frac{1}{r^2} \, \del_{\phi}^2, \\ 
 &\text{and} \quad \Lap = D_{ss} + D_{\phi\phi}.
\end{aligned}
\end{equation} 
The functions for curvature ($\kappa_s$, $\kappa_\phi$), tension ($\tau_s$, $\tau_\phi$) and geometric properties ($\psi$, $r$) occurring in the stability equations (\ref{eq:DMV_stability}) are properties of the axisymmetric buckled configuration that is to be tested for its stability with respect to non-axisymmetric deflections. They are, thus, known numerically when (\ref{eq:DMV_stability}) is solved.

The stability equations are applicable under the same conditions as the DMV theory \cite{Niordson1985}. These prerequisites are (i) the typical length scale of the deformation is much smaller than the smallest radius of curvature of the reference shape, (ii) the displacements are predominantly normal to the surface and (iii) the stresses due to bending are smaller than the stresses due to stretching. They are satisfied in many shallow shell problems; but also for shells which are closed, as in the present case and, therefore, essentially non-shallow as long as  the ``relevant'' part of the shell, which is subjected to the largest deformation, is shallow. Thus, we can expect the theory to be accurate for small dimples, but not for large dimples where the meridian makes a $180^\circ$ turn at the dimple edge. The dimple size at secondary buckling varies with $\tilde E_B$, and sufficiently small dimples are found for small $\tilde E_B$.

In the following, we are searching for non-axisymmetric solutions in the form of wrinkles. They are captured by an ansatz
\begin{equation}
\begin{aligned}
 w(s, \phi) &= W(s) \cos(n \phi), \,\,
 \sphi(s, \phi) = \Phi(s) \cos(n \phi),
\end{aligned}
\label{eq:DMV_ansatz}
\end{equation} 
where $n$ is the number of wrinkles in circumferential direction and $W$ and $\Phi$ are, as in section \ref{sec:wrinkling_crit_curved_plate}, amplitude functions. Inserting this ansatz into (\ref{eq:DMV_stability}) results in ordinary differential equations for the amplitude functions; the only change in the equations is $\del_{\phi}^2 \rightarrow -n^2$.

The boundary conditions for the amplitude functions are similar to the ones in section \ref{sec:wrinkling_crit_curved_plate}: the normal displacement and the additional stresses shall approach 0 far outside the deformed region. In the present case, this means that we choose a sufficiently large $s_\text{max}$, where we require $W(s_\text{max}) = W'(s_\text{max}) = 0$ and $\Phi(s_\text{max}) = \Phi'(s_\text{max}) = 0$. At the point $s = 0$, there are, in principle, no boundary conditions at all, since this is not a point on a boundary. However, the point $s = 0$ causes problems, since some terms in the differential equations diverge here, for example the $1/r$ terms in (\ref{eq:Dss_etc}). This problem will be circumvented by choosing a suitable discretisation scheme (see appendix \ref{app:discretisation}). Considering the ansatz (\ref{eq:DMV_ansatz}), we have to enforce $W(0) = \Phi(0) = 0$ at least, in order to obtain continuous fields of displacement and stress potential.

\subsection{Results for the secondary buckling transition}

\begin{figure}
  \centerline{\includegraphics[width=80mm]{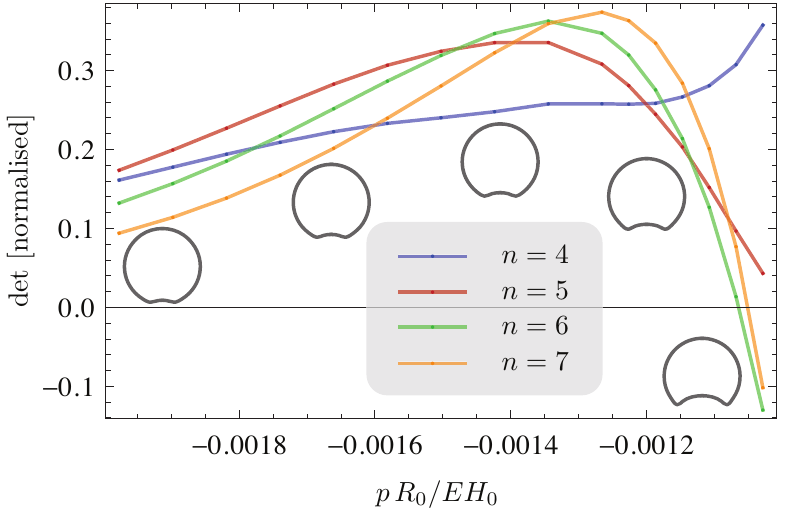}}
  \caption{(Colour online) Normalised determinant of the coefficient matrix of  the discretised stability equations for different wrinkle numbers $n$ for successively deflated, axisymmetric buckled shapes. For $n=6$, the root of the determinant occurs for the smallest deformation. The elastic moduli of the capsule are $\tilde E_B = 10^{-5}$ and $\nu=1/3$.}
\label{fig:determinant}
\end{figure}

The discretisation that we use to transform the differential equations into a linear system is documented in appendix \ref{app:discretisation}. The solution is then represented by the values $W^{(i)} = W\big({s}^{(i)}\big)$ and $\Phi^{(i)} = \Phi\big({s}^{(i)}\big)$ of the amplitude functions evaluated on the grid ${s}^{(i)} = i \cdot h$, with $0\leq i \leq N$ and a step size $h=s_\text{max}/N$.

Since the linear system is homogeneous, a non-trivial solution only exists if the determinant of the coefficient matrix vanishes. Thus, the stability equations turn out to be similar to an eigenvalue problem, as in the curved plate model, where the eigenvalue is ``hidden'' in the axisymmetric buckled shape. Plotting the determinant along the axisymmetric buckled branch, for example as a function of the pressure $p$, the determinant has a root at a critical pressure which is the onset of the wrinkling instability, see fig.\ \ref{fig:determinant}. A similar procedure has been successfully applied to the unsymmetrical buckling of shallow shells \cite{Huang1964,Bushnell1967}. We use the pressure as the parameter to run through the branch of deflated shapes, because it directly enters the shape equations as a Lagrange parameter. The branch consists of shapes which are unstable in experiments with prescribed pressure; however, these shapes are accessible in volume controlled experiments. The first root of the determinant functions occurs, in the example of fig.\ \ref{fig:determinant}, for a wrinkle number $n=6$ at approximately $p = -0.00107 EH_0/R_0$, which corresponds to a volume reduction of $5.5\%$. These are the results for the critical wrinkle number, critical pressure and critical volume difference for the secondary buckling for the specific capsule of fig.\ \ref{fig:determinant}.

\begin{figure}
  \centerline{\includegraphics[width=80mm]{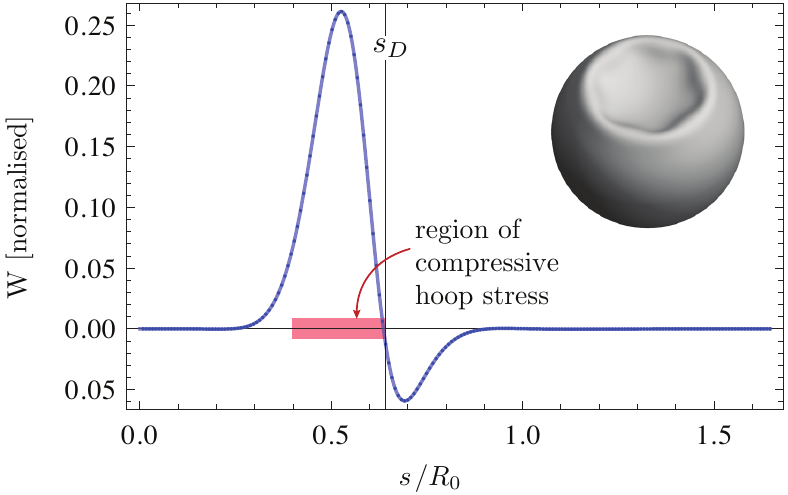}}
  \caption{(Colour online) Solution for the wrinkle amplitude $W$ for a capsule with $\tilde E_B = 10^{-5}$ and $\nu=1/3$ at the secondary buckling transition. The vertical line marks $s_D$, the position of the dimple edge where the curvature $\kappa_s$ is at its maximum. The inset shows a three-dimensional view when this wrinkle profile is added as a normal displacement to the axisymmetric buckled shape, with wrinkle number $n=6$.}
\label{fig:W}
\end{figure}

At the critical volume, the stability equations have a non-trivial solution, which can easily be found by standard methods for linear systems. The resulting wrinkle amplitude $W(s)$, as the interpolation of the $W^{(i)}$ points, is plotted in fig.\ \ref{fig:W} for the same capsule as used for fig.\ \ref{fig:determinant}. In comparison with the curved plate model we find a qualitative agreement (cf.\ figs.\ \ref{fig:krumm_results} and \ref{fig:wrinkles_curved_plate}): The wrinkle amplitude has a prominent peak which is centred in the region of compressive hoop stress, and decays rapidly outside this region. In contrast to the curved plate model, the wrinkle amplitude is not symmetric with respect to its maximum; instead we observe an overshoot only in the outer region, and not towards the centre of the dimple. The most obvious difference in the three-dimensional views is the different wrinkle number, which will be discussed later.

Repeating the above procedure to calculate the critical volume of the secondary buckling for different $\tilde E_B$, we obtain a further line for the phase diagram fig.\ \ref{fig:phase_diag} (orange data points). For sufficiently small bending stiffness, $\tilde E_B < 10^{-5}$ in the present case, the data points can be fitted with a power law
\begin{equation}
 \left. \frac{\Delta V_\text{2nd}}{V_0} \right|_\text{DMV} = 
   (19000\pm900) \, \tilde E_B^{1.110\pm0.004}. 
\end{equation} 
and are close to the previously generated secondary buckling line according to the curved plate model, see fig.\ \ref{fig:phase_diag}. Only for too large bending stiffnesses, the data points deviate from the power law fit and the previous line. This is due to the violation of the assumption that the characteristic length of the deformation is small compared to the radius of curvature, on which the DMV theory is based, but which is not justified for large dimples.


\section{Discussion of results}
\subsection{The complete phase diagram}

\begin{figure*}[t]
  \centerline{\includegraphics[width=165mm]{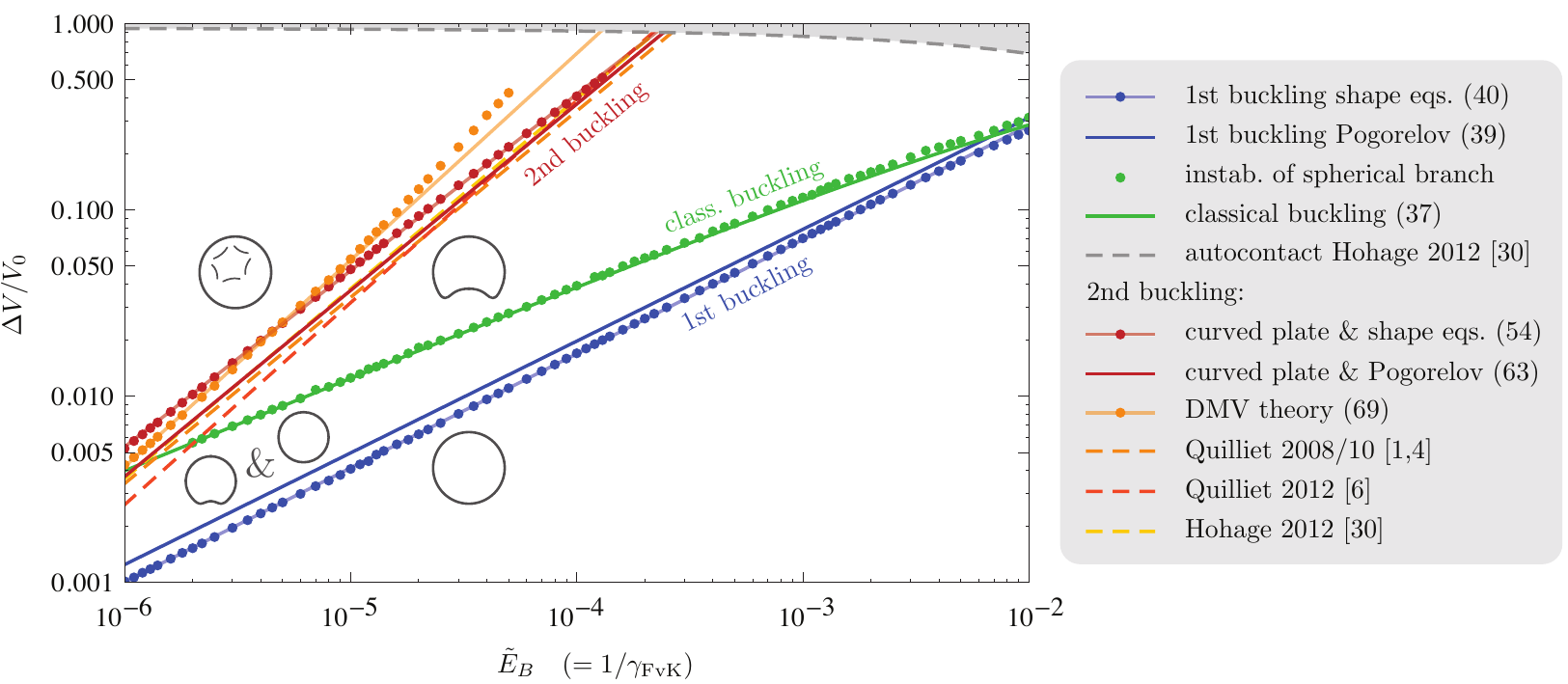}}
  \caption{(Colour online) Phase diagram of deflated spherical shells with Poisson ratio $\nu=1/3$ and varying bending stiffness $\tilde E_B$ (which is related to the F\"oppl-von-K\'arm\'an-number $\gamma_\text{FvK} = 1/\tilde E_B$). Dots represent results derived from numerical solutions of the axisymmetric shape equations, continuous semi-transparent lines (blue, red, orange) fits to those data points, solid lines (blue, green, red) analytic results, and dashed lines results from numerical simulations. The legend to the right indicates the appropriate equation numbers or literature references.}
\label{fig:phase_diag}
\end{figure*}

The complete phase diagram in the $\Delta V$-$\tilde E_B$-plane, i.e., for buckling under volume control containing all transition lines derived by the various models discussed in this paper is shown in fig.\ \ref{fig:phase_diag}. In addition, literature results for the secondary buckling transition, which are all based on numerical simulations, are shown \cite{Quilliet2008,Quilliet2012,Hohage2012}. For completeness, the line of auto-contact, where opposite sides of extremely deflated shells touch each other, is also shown \cite{Hohage2012}.

We will discuss the deflation behaviour of a spherical shell under volume control by following an imaginary vertical line in this phase diagram. Starting at the bottom, at small volume difference, there is only one possible shape for the shell: it is spherical, with a radius smaller than the initial radius.

Upon deflation, the shell will cross the critical volume of first buckling, $\Delta V_\text{1st}$ (blue in fig.\ \ref{fig:phase_diag}). From there on, there are two possible shell configurations: a spherical shape and an axisymmetric buckled shape, of which the axisymmetric buckled shape is  stable and the spherical shape only metastable. Thus, without external perturbations, the shell will remain spherical. However, it is also possible to indent the shell manually, and the dimple will remain on the shell. Thermal fluctuations or imperfections in the geometry or material, as they are inherent to real shells, may also cause a sudden transition to the dimpled shape, although the spherical shape is theoretically (meta)stable.

Deflating the shell further, we will cross the line of classical buckling (green in fig.\ \ref{fig:phase_diag}). At this critical volume $\Delta V_\text{cb}$, the spherical configuration becomes unstable. Thus, if this volume difference is exceeded, the shell must buckle. In the range of $\tilde E_B$ investigated here, the buckled shape is axisymmetric at first; but extrapolating the lines to smaller $\tilde E_B$ suggests that this will not be the case for very small $\tilde E_B$. The transition from spherical to axisymmetric buckled shape is, in any case, a discontinuous abrupt  transition as the energy diagram (fig.\ \ref{fig:bifdiag_U_V}) shows: the dimple will have a finite size when it is formed; shapes with infinitesimal dimples are energy maxima lying above the energy of the spherical branch and represent a possible transition state at the discontinuous transition \cite{Knoche2011}.

\begin{figure}[b]
  \centerline{\includegraphics[width=80mm]{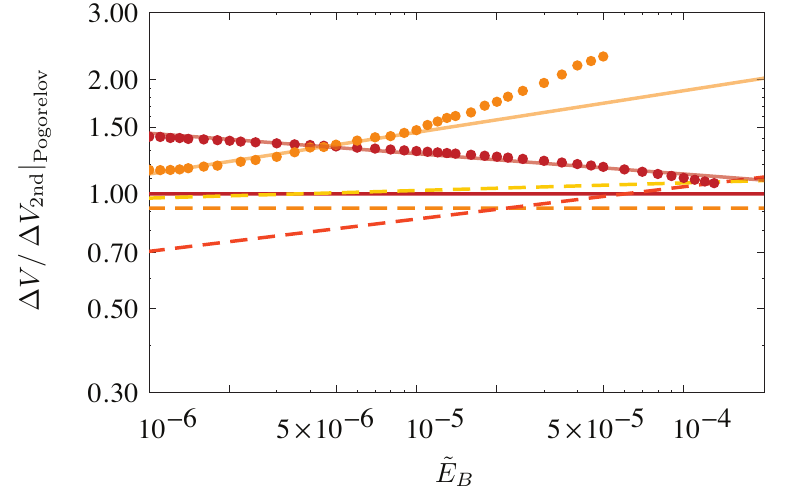}}
  \caption{(Colour online) The secondary buckling lines, normalised by the secondary buckling volume of the Pogorelov model. For the legend, see fig.\ \ref{fig:phase_diag}.}
\label{fig:secbuck_lines}
\end{figure}

The axisymmetric buckled shape will lose its stability with respect to non-axisymmetric deformations when we cross the line of secondary buckling at $\Delta V_\text{2nd}$. Due to a region of strong hoop compression in the inner neighbourhood of the dimple edge, circumferential wrinkles will appear in this region. The circumferential fibres release the strong compression by buckling out of their symmetric shape, similar to a straight rod in the case of Euler buckling \cite{Landau1986}, trading compression energy for additional bending energy. We calculated the line of this secondary buckling transition via three different routes: (i) deriving a  wrinkling criterion for a curved rectangular plate and applying this criterion to the numerical results of the axisymmetric shape equations, (ii) applying the same criterion to the approximate analytical results in the framework of the  Pogorelov model, and (iii) by a  linear stability analysis of the full form in the framework of the DMV shell theory. The results of all three approaches agree well within the regimes of their respective validity. Literature results for the secondary buckling line, based on simulations with the program surface evolver \cite{Quilliet2008,Quilliet2012} or a spring-bead model for a triangulated sphere \cite{Hohage2012}, are also in good agreement with our findings. They are plotted by their power laws
\begin{align}
\Delta V_\text{2nd}/V_0 &= 3400 \, \tilde E_B 
  && \text{from ref.\ \cite{Quilliet2008}} \\
\Delta V_\text{2nd}/V_0 &= 8470 \, \tilde E_B^{1.085} 
   && \text{from ref.\ \cite{Quilliet2012}} \\
\Delta V_\text{2nd}/V_0 &= 4764 \, \tilde E_B^{1.020} 
  && \text{from ref.\ \cite{Hohage2012}},
\end{align}
as obtained from fitting the simulation results. In fact, the different secondary buckling lines are so close that some of them cannot be distinguished. Figure \ref{fig:secbuck_lines} shows a detail plot, in which the different lines have been ``normalised'' by the line of the Pogorelov model (\ref{eq:DeltaV2nd_Pogorelov}) in the sense of plotting
\begin{equation}
 \frac{ \left.{\Delta V_\text{2nd}}\right|_i }
   { \left.{\Delta V_\text{2nd}}\right|_\text{Pog} }
\end{equation} 
for each line $i$ as a function of $\tilde E_B$. This plot  shows that the secondary buckling volume differences obtained in numerical simulations (dashed lines) are typically smaller than our results, which might be due to the fact that the simulated shells are imperfect because of the triangulation. Furthermore, fig.\ \ref{fig:secbuck_lines} makes clear that the stability analysis of the full form in DMV theory (orange data points and fit) is in very good agreement with the linear $\tilde E_B$ dependence from the Pogorelov model for small $\tilde E_B$, but deviates for larger $\tilde E_B$. The reason is, as mentioned before, that for large $\tilde E_B$ the dimple grows too large before the secondary buckling, so that the DMV theory is inaccurate. Indeed, fitting only the seven data points with smallest bending stiffness $\tilde E_B < 2\cdot 10^{-6}$ results in $\left.\Delta V_\text{2nd}\right|_\text{DMV} \propto \tilde E_B^{1.05}$ with an exponent very close to the analytic exponent $1$.

\subsection{The number of wrinkles in secondary buckling}

\begin{figure}
  \centerline{\includegraphics[width=80mm]{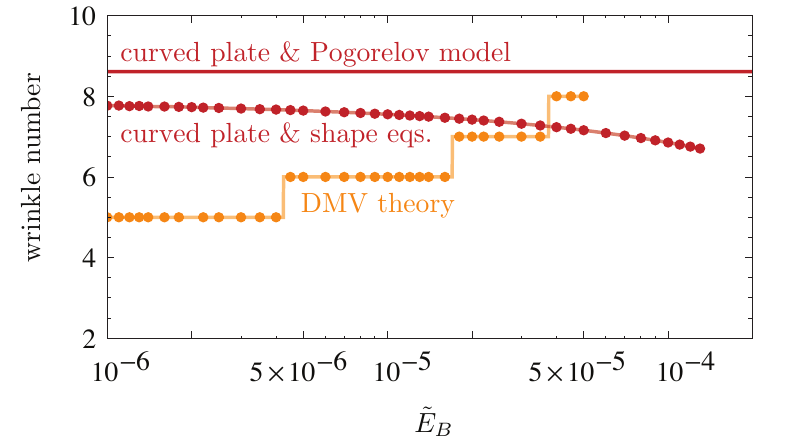}}
  \caption{(Colour online) Number of wrinkles as a function of the bending stiffness at the onset of secondary buckling. The red data points represent the result (\ref{eq:lambda_crit}) from the curved plate model applied to the shape equations and should be rounded to an integer. The red line shows the constant wrinkle number \Cr{$n=8.6$} as obtained from the eq.\ (\ref{eq:lambda_crit}) applied to Pogorelov model. The orange data points represent the results from the DMV theory. 
}
\label{fig:wrinkle_number}
\end{figure}

Although the critical volume of the secondary buckling transition agrees fairly well in all models, there are substantial differences in the predicted number of wrinkles at the onset of secondary buckling, see fig.\ \ref{fig:wrinkle_number}. In deriving the wrinkling criterion (\ref{eq:tau_crit}) for the curved plate we also obtained the unstable wrinkling wavelength at the onset of secondary buckling in eq.\ (\ref{eq:lambda_crit}). Applying this result to the Pogorelov model, we have seen that the wrinkle number is fixed at \Cr{$8.6$}, see eq.\ (\ref{eq:Pogorelov_wrinkle_number}) and the red horizontal line in fig.\ \ref{fig:wrinkle_number}. Applying the  curved plate result (\ref{eq:lambda_crit}) to the shape equations (red data points), we observe a slight decrease in the wrinkle number for increasing bending stiffness. This is contrary to the wrinkle number in the DMV theory (orange data points), which slightly increases with increasing bending stiffness.

The question whether the wrinkle number $n$ at the onset of the secondary buckling should depend on $\tilde E_B$ is difficult to answer from intuition, because there are two opposing effects at work: On the one hand, we expect the wrinkle wavelength $\lambda$ to become smaller for smaller bending stiffness; but on the other hand, the dimple size $r_D$ at the onset of the secondary buckling also becomes smaller (see the phase diagram, fig.~\ref{fig:phase_diag}). The number of wrinkles is $n = 2 \pi r_D/\lambda$, and both effects can cancel out.

Let us investigate the scaling laws, which provide more insight, also into the evolution of the number of wrinkles beyond the onset of secondary buckling. In the simple scaling argument of section \ref{sec:wrinkling_crit_curved_plate} we used that the wrinkle wavelength is given by the width of the compressive part of the stress parabola, $\lambda \sim 1/\sqrt{a_p}$. In the Pogorelov model we saw that the coefficient $a_p$ of the stress parabola scales with the reduced bending stiffness, but is independent of the volume reduction, see (\ref{eq:a_p_Pog}). This leads to the scaling
\begin{equation}
   \lambda \sim R_0 \tilde E_B^{1/4}\sim \sqrt{H_0 R_0},
\end{equation}
which is consistent with previous findings \cite{Quilliet2012,
  Vella2011}. From the geometrical relations (\ref{eq:r_D}) and
(\ref{eq:DeltaV}) it follows that the radius of the dimple is related to the
volume difference by $r_D \sim R_0 (\Delta V/V_0)^{1/4}$.

This results in a  wrinkle number scaling 
\begin{equation} 
  \label{eq:n(EB, DelatV)}
 n  \sim r_D/\lambda \sim \tilde E_B^{-1/4} (\Delta V/V_0)^{1/4}.
\end{equation}
When we control the bending stiffness and volume reduction independently, we find that for fixed volume difference, the wrinkle number decreases with increasing bending stiffness which is in accordance with simulation results \cite{Quilliet2008, Quilliet2012}. Following the deflation of a single capsule with fixed $\tilde E_B$, we find that the wrinkle number should increase with $\Delta V$, which is in accordance with the simulations in ref.\ \cite{Quilliet2008}; in ref.\ \cite{Quilliet2012} an initial decrease of the wrinkle number just after the onset (followed by an increase) is reported that is not captured by this scaling law.

However, when we investigate the wrinkle number directly at the onset of secondary buckling, the volume difference is determined by the bending stiffness and scales linearly with $\tilde E_B$, see (\ref{eq:DeltaV2nd_Pogorelov}) and the phase diagram fig.\ \ref{fig:phase_diag}. Inserting $\Delta V/V_0 \sim \tilde E_B$ into (\ref{eq:n(EB, DelatV)}) shows that the number of wrinkles \emph{at the onset} of secondary buckling does not scale with the bending stiffness. If there are significant variations of the wrinkle number at the onset, the dependence on $\tilde E_B$ comes from more subtle effects. This possibly explains why our different approximate models, which all have their flaws, predict slightly different results.

Comparison with previous simulation results is difficult since most results are given for highly deflated shells, and not at the onset of secondary buckling. In ref.~\cite{Quilliet2008}, four different deflation trajectories are reported, for $\tilde E_B \approx 3 \cdot 10^{-5}$ to $4 \cdot 10^{-4}$. They all exhibit $5$ wrinkles at the onset, except the one with smallest $\tilde E_B$, which starts to wrinkle with $6$ wrinkles. This affirms that the wrinkle number at the onset is quite unaffected by the bending stiffness. In other simulations based on triangulated surface models \cite{Hohage2012, Wischnewski2013} we also observed that the number of wrinkles depends sensitively on the triangulation, specifically on the position and type of disclinations.


\section{Conclusions}

We investigated the deformation  behaviour of spherical elastic shells upon deflation under volume control and developed a complete theory for the generic sequence of equilibrium shapes. For small volume changes, the shell remains in a spherical shape, then jumps to an axisymmetric buckled configuration in a primary buckling transition, and finally undergoes a secondary buckling transition where the dimple loses its axisymmetry. These three shapes represent the possible stable equilibrium states and are, thus, practically most relevant, although many other metastable shapes exist, as shown earlier by a bifurcation analysis of  axisymmetric shapes \cite{Knoche2011}. We obtained quantitative results for the critical volumes of both buckling transitions using  different approaches based on elasticity theory of shells. Our results  are in mutual agreement and in agreement with previous literature results and are summarised in the shape or phase diagram in fig.\ \ref{fig:phase_diag} in the plane of reduced volume difference and reduced bending rigidity $\tilde E_B$ (equal to the inverse  F\"oppl-von-K\'arm\'an-number).

The axisymmetric buckling, out of the spherical shape, occurs at a volume difference somewhere between the first buckling line, $\Delta V_\text{1st} \sim \tilde E_B^{3/5}$, and the classical buckling line, $\Delta V_\text{cb} \sim \tilde E_B^{1/2}$. The first buckling line is defined as the volume difference (in dependence on the bending stiffness) where the elastic energies of the spherical solution branch and the axisymmetric buckled branch cross, so that the axisymmetric buckled shapes are energetically favourable for larger volume differences \cite{Knoche2011}. The classical buckling line is defined by the onset of instability of the spherical shapes. We investigated two different models for the axisymmetric buckling: Firstly, shape equations based on nonlinear shell theory which have to be solved numerically, and secondly, an analytic model which has been introduced  by Pogorelov, but which has been rarely used to derive practical and quantitative  results so far.

The main focus of this paper was on the secondary buckling transition, because this phenomenon is lacking a theoretical explanation, but has been observed in experiments or simulations. The dimple's loss of axisymmetry occurs by wrinkling along the inner side of the dimple edge. The physical explanation for this instability is a hoop stress which is highly compressive in this region, a feature that is well captured by the nonlinear shape equations and the analytic Pogorelov model. An expression for the critical compressive tension was derived in a simplified geometry and stress state, where the midsurface was approximated as a curved plate with a height profile in form of a cubic parabola and a parabolic stress profile to approximate the locally compressive hoop stress. Application of this secondary buckling criterion to the axisymmetric solutions of the shape equations and Pogorelov model gave a line for the secondary buckling transition in the phase diagram, $\Delta V_\text{2nd} \sim \tilde E_B$. The results were verified by a linear stability analysis in the framework of nonlinear DMV shell theory, where the full geometry and state of stress was taken into account. For small $\tilde E_B$, where the dimples are quite small at the secondary buckling transition and the DMV theory is therefore accurate, our previous results were confirmed. Our results are also in good agreement with numerical simulations found in the literature.

Finally, our analysis showed that the primary buckling transition is discontinuous, i.e., we find metastability above and below the transition volume $V_\text{1st}$ and the buckled state has a dimple of finite size, whereas the secondary buckling transition is continuous with a wrinkling amplitude which becomes arbitrarily small at the transition.

We also obtain results for the wrinkle number $n$ at the onset of secondary buckling and beyond. Scaling arguments give $n \sim \tilde E_B^{-1/4} \Delta V^{1/4}$, see eq.\ (\ref{eq:n(EB, DelatV)}), for the number of wrinkles. At the onset of secondary buckling, where $\Delta V_\text{2nd} \sim \tilde E_B$, the  wrinkle number becomes approximately independent of the reduced bending rigidity in accordance with our more detailed analysis, see fig.\ \ref{fig:wrinkle_number}. Beyond the onset of secondary buckling the wrinkle number increases $n \sim \Delta V^{1/4}$ in accordance with numerical simulations in the literature. 

There are still open questions on the secondary buckling of spherical shells. When the pressure inside the shell is prescribed, rather than the volume, large parts of the axisymmetric buckled branch become unstable, and the capsule is likely to buckle through, until opposite sides are in contact \cite{Knoche2011}. In these self-contacting shapes, the secondary buckling may also occur, but in modified form. Furthermore, we currently cannot calculate wrinkled shapes beyond the secondary buckling threshold. Thus, in this regime a quantitative result for the wrinkle number beyond the scaling result (\ref{eq:n(EB, DelatV)})
is still missing.

\begin{acknowledgments}
  We thank Jens Hohage and Christian Wischnewski for their numerical
  simulation results.
\end{acknowledgments}

\appendix
\numberwithin{equation}{section}

\section{Analytic solution of the variational problem}
\label{app:sol_variation}

The analytic solution of Pogorelov's variational problem proceeds as follows. The coordinate range $[0, \infty)$ is divided into two parts $I_1 = [0, \sigma)$ and $I_2 = [\sigma,\infty)$. On the interval $I_1$, the function $\bar w$ is of the order of unity (since its starting value is 1) and it shall have a root at $\bar s_0 = \sigma$. From there on, $\bar w$ is assumed to stay small, that is, we can neglect $\bar w^2$ as compared to $\bar w$ on  $I_2$, which will simplify the constraint (\ref{eq:constraint_nd}). On $I_1$, Pogorelov argues that $\bar w'$ should be approximately constant, because the curvature $\kappa_s \propto \bar w'$ has a maximum at the dimple edge and therefore varies little in its vicinity. With this simplification and the boundary conditions (\ref{eq:BC_nd}) we have
\begin{equation}
 \bar w_1(\bar s_0) = (\bar s_0 - \sigma)/\sigma,
 \label{eq:v_1}
\end{equation} 
and the constraint (\ref{eq:constraint_nd}) and boundary conditions further dictate
\begin{equation}
 \bar u_1 (\bar s_0) = -\frac{1}{2\sigma} (\bar s_0 - \sigma)^2 
  - \frac{1}{6\sigma^2} (\bar s_0 - \sigma)^3 + \frac{\sigma}{3}
 \label{eq:u_1}
\end{equation} 
as the solution on $I_1$. On $I_2$, the constraint simplifies to $\bar w_2 = -\bar u_2'$, which can be inserted directly into the energy functional (\ref{eq:J_def}). The complete functional, on $I_1$ and $I_2$, reduces with these two simplifications to
\begin{align}
 J &= \int_0^\sigma \diff \bar s_0 \left\{ \bar w_1'^2 + \bar u_1^2 \right\}
      +\int_\sigma^\infty \diff \bar s_0 \left\{ \bar w_2'^2 + \bar u_2^2 \right\} 
      \nonumber \\
   &= \frac{1}{\sigma} + \frac{17}{315} \sigma^3
      +\int_\sigma^\infty \diff \bar s_0 \left\{ \bar u_2''^2 + \bar u_2^2 \right\}. 
   \label{eq:J(sigma,u)}
\end{align}
With the Ansatz  on $I_1$, the functions $\bar u_1$ and $\bar w_1$ are fixed by (\ref{eq:v_1}) and (\ref{eq:u_1}), respectively. A variation  is only possible by varying the parameter $\sigma$. On $I_2$, the function $\bar u_2$ can be subjected to arbitrary variations which respect the boundary conditions. At $\bar s_0 = \sigma$, the boundary condition is given by the continuity condition $\bar u_2(\sigma) = \bar u_1(\sigma) = \sigma/3$. To find the minimum of (\ref{eq:J(sigma,u)}), we first keep $\sigma$ fixed and variate with respect to $\bar u_2$. The solution will depend still on $\sigma$, and we then minimise with respect to $\sigma$.

Requiring a vanishing variation $\delta J[\bar u_2]=0$ results in the differential equation
\begin{equation}
 \bar u_2'''' + \bar u_2 = 0.
\end{equation} 
It can be solved with an exponential ansatz, and the solution which conforms the boundary conditions is given by
\begin{align}
 \bar u_2(\bar s_0) = -\frac{\sigma}{3\sqrt{2}}
\left( \omega_1 e^{\omega_1(\bar s_0 - \sigma)} + 
     \omega_2 e^{\omega_2(\bar s_0 - \sigma)} \right)
 \label{eq:u_2}
\end{align} 
with $\omega_1 = -(1-i)/\sqrt{2}$ and $\omega_2 = -(1+i)/\sqrt{2}$. From the constraint $\bar w_2 = -\bar u'$ it follows immediately that
\begin{equation}
 \bar w_2 = \frac{i \sigma}{3\sqrt{2}} 
  \left( -e^{\omega_1(\bar s_0 - \sigma)} + e^{\omega_2(\bar s_0 - \sigma)} \right).
 \label{eq:v_2}
\end{equation} 
With eqs.\ (\ref{eq:v_1}), (\ref{eq:u_1}), (\ref{eq:u_2}) and (\ref{eq:v_2}), the complete solution is determined,
\begin{align}
 \bar u(\bar s_0) = 
 \begin{cases}
  \bar u_1(\bar s_0), & 0\leq \bar s_0 < \sigma \\
  \bar u_2(\bar s_0), & \bar s_0 \geq \sigma
 \end{cases} \\ 
 \bar w(\bar s_0) = 
 \begin{cases}
  \bar w_1(\bar s_0), & 0\leq \bar s_0 < \sigma \\
  \bar w_2(\bar s_0), & \bar s_0 \geq \sigma
 \end{cases}.
\end{align} 

Evaluating the functional $J$ for this solution, we get
\begin{equation}
 J(\sigma) = \frac{1}{\sigma} + \frac{\sqrt{2}}{9} \sigma^2
      + \frac{17}{315} \sigma^3.
\end{equation} 
We can now perform the final minimisation with respect to $\sigma$, which gives a numerical value of 
\begin{equation}
 \sigma_{\rm min} = 1.24667 \quad \text{and} \quad J_{\rm min} = 1.15092.
 \label{eq:sigma_J_min}
\end{equation} 


\section{Wrinkling of curved plates}
\label{app:wrinkling}

In this appendix, we show how the stability equations of shallow shells  (\ref{eq:stability_1}) and (\ref{eq:stability_2}) can be derived from an energy functional. These energy considerations also serve to classify the secondary buckling as a \emph{continuous} transition. 

We consider a shallow shell above the $x$-$y$-plane with curvatures $\kappa_x$ and $\kappa_y$. It is subjected to in-plane stresses $\tau_x$, $\tau_y$ and $\tau_{xy}$, and an external normal load $p$. We do not make any assumptions on the in-plane-stresses; they may be generated from linear or nonlinear elasticity,  thermal stresses, residual stresses from plastic deformations or inhomogeneous growth, or whatever we can think about -- as long as they satisfy certain equilibrium conditions as presented below. This general framework is necessary in order to capture the specific case analysed in section \ref{sec:secbuck_wrinkling}, because the parabolic stress state does not satisfy the compatibility conditions of linear elasticity.

When we add small displacements $u$, $v$, $w$ (in $x$, $y$ and $z$ direction, respectively) to the given state of the shallow shell, its elastic energy will change. If we can find a displacement field which lowers the elastic energy, the given state is unstable. The fields $u$ and $v$ are not to be confused with the fields in the Pogorelov model; the notation in this appendix is chosen as in ref.\ \cite{Ventsel2001}. 

The strains and bending strains induced by the displacement field are given by 
\begin{align}
 \epsilon_x &= \frac{\del u}{\del x} - \kappa_x w + \frac{1}{2}\left(\frac{\del w}{\del x}\right)^2, 
 & K_x &= \frac{\del^2 w}{\del x^2} \\
 \epsilon_y &= \frac{\del v}{\del y} - \kappa_y w + \frac{1}{2}\left(\frac{\del w}{\del y}\right)^2, 
 & K_y &= \frac{\del^2 w}{\del y^2} \\
 \epsilon_{xy} &= \frac{1}{2} \left(\frac{\del u}{\del y} + \frac{\del v}{\del x} + \frac{\del w}{\del x}\frac{\del w}{\del y}\right),
 & K_{xy} &= \frac{\del^2 w}{\del x \del y}
\end{align} 
(cf.\ ref.\ \cite{Ventsel2001}, p.\ 523). The total energy variation can be written as a surface integral
\begin{equation}
 \Delta W = \int \diff A \left\{ w_\text{stretch} + w_\text{bend} + w_\text{ext} \right\}
\end{equation} 
over the energy densities
\begin{align}
 w_\text{stretch} &= \frac{1}{2} \frac{EH_0}{1-\nu^2} \left( \epsilon_x^2 + 2\nu \epsilon_x \epsilon_y + \epsilon_y^2 + 2 (1-\nu) \epsilon_{xy}^2 \right) \nonumber \\
 & + \tau_x \epsilon_x + 2 \tau_{xy} \epsilon_{xy} + \tau_y \epsilon_y\\
 w_\text{bend} &= \frac{1}{2} E_B \left( K_x^2 + 2\nu K_x K_y + 2 (1-\nu) K_{xy}^2 \right) \\
 w_\text{ext} &= p w.
\end{align}
We sort the energy functional by different orders of the displacement fields, $\Delta W = \Delta W^{(1)} + \Delta W^{(2)} + \Delta W^{(3)} + \Delta W^{(4)}$. 

The \emph{first variation} contains all terms linear in $u$, $v$ and $w$ and reads
\begin{multline}
 \Delta W^{(1)} = \int \diff A \big\{ u \left[ -\del_x \tau_x - \del_y \tau_{xy} \right] + v \left[ -\del_y \tau_y - \del_x \tau_{xy} \right] \\
 + w \left[ -\kappa_x \tau_x - \kappa_y \tau_y + p \right] \big\}
 \label{eq:Delta_W_1}
\end{multline}
after integration by parts has been used (we omit the boundary terms). Linear stability requires the contents of the three square brackets to vanish, which gives the ordinary stability equations for membranes (without bending).

The \emph{second variation} contains all quadratic terms, including terms mixing different fields. Integration by parts can be used to write the integrand in a symmetric form. This results in
\begin{equation}
 \Delta W^{(2)} = \int \diff A \left\{ (u, v, w) \hat H (u, v, w)^T \right\}
\label{eq:Delta_W_2}
\end{equation}
with an operator $\hat H$ which is a $3\times3$ matrix that contains differential operators in its components. It is self-adjoint and, therefore, has an eigenbasis. Thus, $\hat H$ is positive definite if all its eigenvalues are positive. When the lowest eigenvalue falls below zero, $\hat H$ is not positive definite and, thus, a deformation mode exists which lowers the elastic energy. This is exactly the critical point where wrinkling can occur. We can find the critical point by setting $\hat H (u, v, w)^T = (0, 0, 0)^T$. These three equations are equivalent to
\begin{align}
 E_B \Lap^2 w &= \tau_x \frac{\del^2 w}{\del x^2} + 2 \tau_{xy} \frac{\del^2 w}{\del x \del y} + \tau_y \frac{\del^2 w}{\del y^2} + \kappa_x \tau_x^{(1)} + \kappa_y \tau_y^{(1)} \label{eq:stability_deriv}\\
 0 &= \del_x \tau_x^{(1)} + \del_y \tau_{xy}^{(1)} \label{eq:equil_tau1_1}\\
 0 &= \del_y \tau_y^{(1)} + \del_x \tau_{xy}^{(1)}.\label{eq:equil_tau1_2}
\end{align}
Here, we introduced the additional tensions, which develop due to the strains in linear order,
\begin{align}
 \tau_x^{(1)} &= \frac{EH_0}{1-\nu^2} \left( \epsilon_x^{(1)} + \nu \epsilon_y^{(1)} \right), \label{eq:tau_x_1}
 & \epsilon_x^{(1)} &= \frac{\del u}{\del x} - \kappa_x w \\
 \tau_y^{(1)} &= \frac{EH_0}{1-\nu^2} \left( \epsilon_y^{(1)} + \nu \epsilon_x^{(1)} \right), \label{eq:tau_y_1}
 & \epsilon_y^{(1)} &= \frac{\del v}{\del y} - \kappa_y w  \\
 \tau_{xy}^{(1)} &= \frac{EH_0}{1+\nu} \epsilon_{xy}^{(1)},
 & \epsilon_{xy}^{(1)} &= \frac{1}{2} \left( \frac{\del u}{\del y} + \frac{\del v}{\del x}  \right). \label{eq:tau_xy_1}
\end{align}
Equations (\ref{eq:equil_tau1_1}) and (\ref{eq:equil_tau1_2}) can be solved by introducing the Airy stress function $\sphi$ \cite{Landau1986,Ventsel2001}, defined by the relations $\tau_x^{(1)} = \del_y^2 \sphi$, $\tau_{xy}^{(1)} = -\del_x \del_y \sphi$ and $\tau_y^{(1)} = \del_x^2 \sphi$. Additionally, we must assure that the tensions derived from the stress function are compatible with Hooke's law, (\ref{eq:tau_x_1}) - (\ref{eq:tau_xy_1}). With Hooke's law, the strains can be expressed in terms of the stress function,
\begin{align}
 \frac{1}{EH_0} \left( \frac{\del^2 \sphi}{\del y^2} -\nu \frac{\del^2 \sphi}{\del x^2} \right) &= \epsilon_x^{(1)} = \frac{\del u}{\del x} - \kappa_x w \label{eq:epsilon_x_1} \\ 
 \frac{1}{EH_0} \left( \frac{\del^2 \sphi}{\del x^2} -\nu \frac{\del^2 \sphi}{\del y^2} \right) &= \epsilon_y^{(1)} = \frac{\del v}{\del y} - \kappa_y w \label{eq:epsilon_y_1}\\
 -\frac{1+\nu}{EH_0} \frac{\del^2 \sphi}{\del x \del y} &= \epsilon_{xy}^{(1)} = \frac{1}{2} \left(\frac{\del u}{\del y} + \frac{\del v}{\del x}\right). \label{eq:epsilon_xy_1}
\end{align}
As the strains are derivatives of the displacements, they must satisfy certain integrability conditions. We can find these conditions, also known as \emph{compatibility conditions} \cite{Ventsel2001}, by eliminating the in-plane displacements $u$ and $v$ from (\ref{eq:epsilon_x_1}) - (\ref{eq:epsilon_xy_1}) using the combination $\del_y^2 \epsilon_x^{(1)} + \del_x^2 \epsilon_y^{(1)} - 2 \del_x \del_x \epsilon_{xy}^{(1)}$. With the six equations (\ref{eq:epsilon_x_1}) - (\ref{eq:epsilon_xy_1}) we can derive from this combination
\begin{equation}
 \Lap ^2 \sphi = -EH_0 \left( \kappa_x \del_y^2 w + \kappa_y \del_x^2 w \right),
\end{equation}  
which is (\ref{eq:stability_1}) in the main text. When the Airy stress function is introduced in (\ref{eq:stability_deriv}), we obtain
\begin{equation}
 E_B \Lap^2 w = \tau_x \del_x^2 w + 2 \tau_{xy} \del_x \del_y w + \tau_y \del_y^2 w + \kappa_x \del_y^2 \sphi + \kappa_y \del_x^2 \sphi
\end{equation}
which is (\ref{eq:stability_2}) in the main text. The solution of these stability equations determines the shape of the wrinkles when the instability sets in. As the stability equations are homogeneous, the amplitude of the wrinkles is arbitrary, in a mathematical sense. In practice, we have to take into account higher order terms in the elastic energy for large amplitudes; from these higher order terms we may also obtain the magnitude of the wrinkle amplitude as we will demonstrate below.

The \emph{third variation} of the energy functional shall be calculated for a given wrinkling mode $(w, \sphi)$ which satisfies the stability equations (\ref{eq:stability_deriv}) - (\ref{eq:equil_tau1_2}). Under this assumption, the third order terms in the energy functional can be simplified to
\begin{multline}
 \Delta W^{(3)} = - \frac{1}{2} \int \diff A \; w \cdot \left\{ \frac{\del^2 w}{\del x^2}\frac{\del^2 \sphi}{\del y^2} \right. \\
 \left. - 2 \frac{\del^2 w}{\del x \del y} \frac{\del^2 \sphi}{\del x \del y} + \frac{\del^2 w}{\del y^2} \frac{\del^2 \sphi}{\del x^2} \right\}.
\end{multline} 
In our case discussed in the main text, the solution (\ref{eq:ansatz_stabeq}) is symmetric, that is, inverting the amplitude and shifting the wrinkles by half a wavelength in $x$ direction results in exactly the same shape. Thus, $\Delta W(w, \sphi) = \Delta W(-w, -\sphi)$, and the odd orders vanish: $\Delta W^{(3)} = 0$.

For the \emph{fourth variation} of the energy we obtain
\begin{equation}
 \Delta W^{(4)} = \frac{1}{8} \frac{EH_0}{1-\nu^2} \int \diff A \left[ \left(\frac{\del w}{\del x}\right)^2 + \left(\frac{\del w}{\del y}\right)^2\right]^2. \label{eq:Delta_W_4}
\end{equation} 
Evidently, this term is positive and limits the magnitude of the wrinkle amplitude.

In the following, we will calculate the wrinkle amplitude as a function of the compressive tension for the case discussed in the main text. This case is specified by $\tau_x = -\tau_0(1 - a_p y^2)$, $\kappa_y = a_c y$ and $\tau_y=\tau_{xy}=\kappa_x=p = 0$, which is linearly stable, cf.\ (\ref{eq:Delta_W_1}). From (\ref{eq:Delta_W_2}) we obtain, by introducing the Airy stress function, 
\begin{multline}
 \Delta W^{(2)} = \frac{1}{2} \int \diff A \; w \left\{ E_B \Lap^2 w \right. \\
 \left. + \tau_0 \left(1-a_p y^2\right) \del_x^2 w - a_c y \del_x^2 \sphi \right\}.
\end{multline} 
We checked with our numerical solutions $(w, \sphi)$ that this integral vanishes at the critical point $\tau_0 = \tau_c$ (\ref{eq:tau_crit}). To calculate the energy decrease for $\tau_0 > \tau_c$, we will use the same functions $(w, \sphi)$ which were calculated at the critical point. This is analogous to the first order perturbation theory used in quantum mechanics. Since our plate is infinitely long in $x$-direction, it is appropriate to discuss the energy decrease per length, and so the integrals in $x$-direction over the trigonometric functions are replaced by their respective averages, i.e.\ $\int_0^L \sin^2(kx)/L = 1/2$ etc. Furthermore, we will switch to the non-dimensional quantities introduced in (\ref{eq:nondim}). We finally obtain
\begin{align}
 \Delta \hat U^{(2)} &= \frac{\Delta \hat W^{(2)}}{L}  \\
 &= \frac{1}{2} (\hat \tau_0 - \hat \tau_c) \hat W_0^2 \int_0^{\hat y_\text{max}} \diff \hat y \; \hat w \left\{ -\hat k^2 \left(1-\hat y^2\right) \right\} \hat w. \nonumber
\end{align} 
In this equation, the function $\hat w$ has a fixed amplitude of $\hat w(0) = 1$, and $\hat W_0$ denotes the actual wrinkle amplitude. The integral is negative since $\hat w$ decays rapidly for $\hat y > 1$. For $\hat a_c = 20$, for example, the integral has a numerical value of $-b_2 = -1.0188$.

From (\ref{eq:Delta_W_4}) we obtain analogously, with $\int_0^L \sin^4(kx)/L = 3/8$ and $\int_0^L \cos^2(kx)\sin^2(kx)/L = 1/8$,
\begin{multline}
 \Delta \hat U^{(4)} = \frac{1}{32} \frac{1}{1-\nu^2} \hat W_0^4 \int_0^{\hat y_\text{max}} \diff \hat y \;\left[ 3 \big(\hat k \hat w\big)^4 \right. \\
 \left. + \big(\hat k \hat w \del_{\hat y}\hat w\big)^2 + 3 \big( \del_{\hat y} \hat w\big)^4 \right],
\end{multline} 
where the integral evaluates to $b_4 = 9.34$ for $\hat a_c = 20$.

\begin{figure}
  \centerline{\includegraphics[width=80mm]{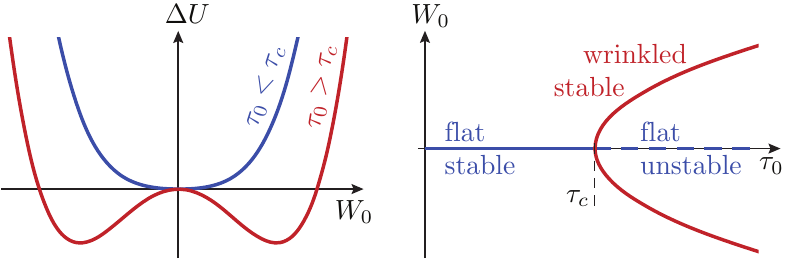}}
  \caption{(Colour online) Bifurcation behaviour for the compressed plate. The elastic energy $\Delta U$ shows, as a function of the wrinkle amplitude $W_0$, either one minimum at $W_0 = 0$ (for $\tau_0 \leq \tau_c$) or two minima and one maximum (for $\tau_0 > \tau_c$). For the equilibrium configuration, which resides in one of the minima, we get a behaviour that changes at $\tau_c$: The flat shape ($W_0=0$) splits at $\tau_c$ into two stable wrinkled shapes, and the flat shape becomes unstable.}
\label{fig:bif_platte}
\end{figure}

So in total we have an elastic energy
\begin{equation}
 \Delta \hat U = -\frac{1}{2} b_2 (\hat \tau_0 - \hat \tau_c) \hat W_0^2 + \frac{1}{32} b_4 \frac{1}{1-\nu^2} \hat W_0^4.
\end{equation} 
The optimal wrinkle amplitude is obtained by minimising this function with respect to $\hat W_0$, which gives
\begin{equation}
 \hat W_0 = \pm \sqrt{\frac{8b_2}{b_4} (1-\nu^2) (\hat \tau_0 - \hat \tau_c)}
\end{equation} 
for $\hat \tau_0 > \hat \tau_c$. The energy function and the position of the (un)stable extrema can nicely be visualised, see fig.\ \ref{fig:bif_platte}. 

The wrinkling of the shallow shell thus represents a \emph{supercritical pitchfork bifurcation}. In contrast to the first buckling transition, the wrinkling is hence a continuous transition.

\section{Stability equations of shells of revolution}
\label{app:stab_DMV}

In this appendix, the derivation of the Cartesian stability equations of the previous section is transferred to axisymmetric shells. The procedure is analogous, but mathematically more involved, and we present only the main steps of the derivation here.

The reference shape for our energy considerations is the axisymmetric buckled shape, characterised by its geometry (the functions $r$, $z$, $\psi$, $\kappa_s$ and $\kappa_\phi$, see section \ref{sec:nonlinear_shape_equations}) and its stress state ($\tau_s$, $\tau_\phi$, $m_s$ and $m_\phi$). This shape is perturbed by displacement fields $u(s, \phi)$, $v(s, \phi)$ and $w(s, \phi)$ in meridional, circumferential and normal direction, respectively. The displacements induce strains according to the strain-displacement relations of the DMV theory \cite{Niordson1985,Ventsel2001}
\begin{align}
 \epsilon_s &= \del_s u - \kappa_s w + \frac{1}{2} (\del_s w)^2 \label{eq:DMV_epsilon_s} \\
 \epsilon_\phi &= \frac{\cos \psi}{r} u + \frac{1}{r} \del_\phi v - \kappa_\phi w + \frac{1}{2 r^2} (\del_\phi w)^2 \\
 \epsilon_{s\phi} &= \frac{1}{2} \left( \frac{1}{r}\del_\phi u - \frac{\cos\psi}{r}v + \del_s v + \frac{1}{r} (\del_s w) (\del_\phi w) \right) \label{eq:DMV_epsilon_sphi}
\end{align}
and bending strains
\begin{align}
 K_s &= \del_s^2 w & \equiv D_{ss} w \\
 K_\phi &= \frac{\cos\psi}{r} \del_s w + \frac{1}{r^2} \del_\phi^2 w &\equiv D_{\phi\phi} w \\
 K_{s\phi} &= \frac{1}{r} \del_s\del_\phi w - \frac{\cos\psi}{r^2} \del_\phi w &\equiv D_{s\phi} w,
\end{align} 
which are justified if the displacements are predominantly normal to the surface, i.e.\ $w$ is larger than $u$ and $v$. The energy change due to these strains is
\begin{equation}
 \Delta W = \int \diff A \left\{ w_\text{stretch} + w_\text{bend} + w_\text{ext} \right\}
\end{equation} 
with the energy densities
\begin{align}
 w_\text{stretch} &= \frac{1}{2} \frac{EH_0}{1-\nu^2} \left( \epsilon_s^2 + 2\nu \epsilon_s \epsilon_\phi + \epsilon_\phi^2 + 2 (1-\nu) \epsilon_{s\phi}^2 \right) \nonumber \\
 & + \tau_s \epsilon_s + \tau_\phi \epsilon_\phi\\
 w_\text{bend} &= \frac{1}{2} E_B \left( K_s^2 + 2\nu K_s K_\phi + 2 (1-\nu) K_{s\phi}^2 \right) \nonumber \\
 & + m_s K_s + m_\phi K_\phi\\
 w_\text{ext} &= p w.
\end{align}

Collecting the first order terms (in $u$, $v$, $w$) and requiring them to vanish for arbitrary displacement fields reproduces the equilibrium equations of the DMV theory. They coincide with (\ref{eq:equil1}) - (\ref{eq:equil3}), except that the effect of the transverse shearing force $q$ is neglected in the meridional force balance (\ref{eq:equil1}), which is typical for the DMV theory \cite{Niordson1985}.

Here we are mainly interested in the second order terms. They can be simplified and symmetrised by integration by parts. The Jacobian determinant in the area element $\diff A = r \diff s \diff \phi$ must be incorporated in the integration by parts, and we obtain as a general rule \cite{Audoly2010}
\begin{equation}
 \int f\cdot(\del_s g) \, \diff A = - \int \frac{1}{r}\,\del_s(r f) \cdot  g \, \diff A,
\end{equation} 
where boundary terms are omitted. As a simplification, we assume that the angle $\psi$ is varying slowly (in comparison to $u$, $v$, $w$ and $r$), so that its derivative $\del_s \psi \approx 0$ can be neglected. This limits the DMV theory to the case where the typical length scale of the non-axisymmetric perturbation is much smaller than the radii of curvature of the axisymmetric buckled shape.

After a quite involved calculation, the energy change in second order can be written as
\begin{equation}
 \Delta W^{(2)} = \int \diff A \left\{ (u, v, w) \hat H (u, v, w)^T \right\}.
\end{equation} 
Following the rationale of appendix \ref{app:wrinkling}, the critical point for the loss of stability of the axisymmetric shape is when the lowest eigenvalue of $\hat H$ is zero. The equations $0=\hat H (u, v, w)^T$ can be recasted into the form
\begin{align}
 0 &= E_B \Lap^2 w - \tau_s D_{ss} w - \tau_\phi D_{\phi\phi} w - \kappa_s \tau_s^{(1)} - \kappa_\phi \tau_\phi^{(1)} \nonumber \\
 0 &= \frac{1}{r} \frac{\del\big(r \tau_s^{(1)}\big)}{\del s} - \frac{\cos\psi}{r} \tau_\phi^{(1)} + \frac{1}{r} \frac{\del \tau_{s\phi}^{(1)}}{\del \phi} \label{eq:DMV_three_stability} \\
 0 &= \frac{1}{r} \frac{\del \tau_{\phi}^{(1)}}{\del \phi} + \frac{1}{r^2}\frac{\del\big(r^2 \tau_{s\phi}^{(1)}\big)}{\del s} \nonumber.
\end{align} 
Here, $\Lap = D_{ss} + D_{\phi\phi}$ as in the main text, and
\begin{equation}
 \begin{aligned}
  \tau_s^{(1)} &= \frac{EH_0}{1-\nu^2} \big(\epsilon_s^{(1)} + \nu \epsilon_\phi^{(1)}\big), 
  &\tau_{s\phi}^{(1)} &= \frac{EH_0}{1+\nu} \epsilon_{s\phi}^{(1)} \\
  \tau_\phi^{(1)} &= \frac{EH_0}{1-\nu^2} \big(\epsilon_\phi^{(1)} + \nu \epsilon_s^{(1)}\big)
 \end{aligned}
\end{equation} 
are the additional tension, which are functions the linearised versions of the strains (\ref{eq:DMV_epsilon_s}) - (\ref{eq:DMV_epsilon_sphi}). The last two equations of (\ref{eq:DMV_three_stability}) are the in-plane equilibrium equations of the general linear membrane theory of shells of revolution \cite{Ventsel2001}. Analogous to the Cartesian case, they are satisfied automatically by using the stress potential $\sphi$, from which the tensions derive as $\tau_s^{(1)} = D_{\phi\phi} \sphi$, $\tau_\phi^{(1)} = D_{ss} \sphi$ and $\tau_{s\phi}^{(1)} = -D_{s\phi} \sphi$. The governing equation for the stress potential is
\begin{equation}
 \Lap^2 \sphi = EH_0 \left( -\kappa_\phi D_{ss} w - \kappa_s D_{\phi\phi} w \right).
\end{equation} 
This equation and the first equation of (\ref{eq:DMV_three_stability}) form the stability equations of axisymmetric shells and are (\ref{eq:DMV_stability}) in the main text.

\section{Discretisation and numerical solution}
\label{app:discretisation}
For the numerical solution of the DMV stability equations, we discretise the differential equations (\ref{eq:DMV_stability}) to obtain a system of linear equations. 

For the nondimensionalisation of the equations, we use the same convention as in section \ref{sec:nonlinear_shape_equations}, i.e.\ we take $R_0$ as the length unit and $EH_0$ as the tension unit. A useful discretisation can be adopted from the literature on the numerical solution of the Poisson equation in polar coordinates \cite{Swarztrauber1973,Lai2001,Lai2002}, because the Poisson equation involves the same problems at $s = 0$ concerning divergences in the Laplacian. So we divide the domain $0 \leq s \leq s_\text{max}$ into $N$ intervals, separated by the points ${s}^{(i)} = i \cdot h$, with $0\leq i \leq N$ and a step size $h=s_\text{max}/N$. The functions $W(s)$ and $\Phi(s)$ are then represented by its values on these sampling points, $W^{(i)} = W\big({s}^{(i)}\big)$ and $\Phi^{(i)} = \Phi\big({s}^{(i)}\big)$.

In the stability equations (\ref{eq:DMV_stability}), derivatives with respect to $s$ up to fourth order occur. In the discretised equations they are approximated by central finite differences,
\begin{align*}
 f'\big({s}^{(i)}\big) &= \frac{-f^{(i-1)} + f^{(i+1)}}{2h} \\
 f''\big({s}^{(i)}\big) &= \frac{f^{(i-1)} - 2f^{(i)} + f^{(i+1)} }{h^2} \\
 f'''\big({s}^{(i)}\big) &= 
   \frac{-f^{(i-2)} + 2f^{(i-1)} - 2f^{(i+1)} + f^{(i+2)} }{2h^3} \\
 f''''\big({s}^{(i)}\big) &= 
  \frac{ f^{(i-2)} - 4f^{(i-1)} + 6f^{(i)} - 4f^{(i+1)} + f^{(i+2)} }{h^4}.
\end{align*}
At the boundaries of the integration region, this involves problems since these formulas are ``overlapping'' the integration region. Thus, to evaluate the third and fourth derivatives at $i=N$, for example, we need to introduce two ``phantom points'' $i=N+1$ and $i=N+2$. That induces four further degrees of freedom in our equations, $W^{(N+1)}$, $W^{(N+2)}$, $\Phi^{(N+1)}$ and $\Phi^{(N+2)}$, and thus necessitates four extra equations: the boundary conditions. As discussed before, we impose a vanishing function value and derivative value at $s_\text{max}$ for both functions $W$ and $\Phi$. In the discretised formulation, this means $W^{(N)} = 0$ and $W^{(N+1)}-W^{(N-1)} = 0$, and the same for $\Phi$.

The boundary $i=0$ is more difficult to handle, since some terms of the stability equations diverge. This problem can be circumvented by transforming the differential equations into a weak form by integrating them in the $(s, \phi)$ space over a disc of radius $\epsilon \rightarrow 0$, i.e.\ integrating over the range $0 \leq s \leq \epsilon$ and $0\leq\phi<2\pi$. In the vicinity of $s = 0$, the axisymmetric solutions satisfy $\kappa_s = \kappa_\phi \equiv \kappa$ and $\tau_s = \tau_\phi \equiv \tau$ \cite{Knoche2011}. Furthermore, $\cos\psi \approx 1$ and thus $r \approx s$, so that the Laplacian reduces to $\Lap = \del_{s}^2 + \frac{1}{s} \del_{s} + \frac{1}{s^2} \del_{\phi}^2$, the usual Laplacian for flat polar coordinates. Thus, the second of the stability equations (\ref{eq:DMV_stability}) reads
\begin{equation}
 0 = \Lap^2 \sphi_1 + EH_0\kappa \Lap w_1.
\end{equation}
Integrating both sides over the disk and using Gauss's theorem to transform the surface integral into a contour integral, we obtain
\begin{multline}
 0 = \int_0^{2\pi} \diff \phi \cos(n \phi_0)
    \left\{ s \Phi''' +  \Phi'' - \frac{n^2+1}{s} \Phi' \right.\\
 \left.\left.+ \frac{2n^2}{s^2} \Phi + 
   EH_0\kappa W' \right\}\right|_{s = \epsilon}.
\end{multline} 
For the other stability equation, we obtain analogously
\begin{multline}
 0 = \int_0^{2\pi} \diff \phi \cos(n \phi) 
 \left\{ E_B \left( s W''' +  W'' - \frac{n^2+1}{s} W' \right.\right.\\
 \left.\left.\left. + \frac{2n^2}{s^2} W \right)
    - \kappa \Phi' - \tau W' \right\}\right|_{s = \epsilon}.
\end{multline} 
As the integral over a full period of the cosine vanishes, these two equations are satisfied when the terms in curly braces do not diverge. This is, for the limit $\epsilon \rightarrow 0$, the case if
\begin{equation}
  W(0) = W'(0) = 0 \quad  \text{and} \quad \Phi(0) = \Phi'(0) = 0. 
 \label{eq:discret_BC_0}
\end{equation}
This has the same form as the boundary conditions at the other end of the integration region; but in this case, it is the expression of the differential equations to be satisfied at ${s}^{(0)}$. Thus, the differential equations (\ref{eq:DMV_stability}) must be imposed only at the points ${s}^{(1)}$, ${s}^{(2)}$, $\dots$, where all terms are regular; at ${s}^{(0)}$ we impose (\ref{eq:discret_BC_0}). By that, we have avoided the problem of diverging terms in the Laplacian.

Hence, we only need one phantom point ${s}^{(-1)}$, at which the function values are fixed by the conditions (\ref{eq:discret_BC_0}) to $W^{(-1)}=W^{(1)}$ and $\Phi^{(-1)}=\Phi^{(1)}$. For the solution of the linear system, we can even spare this phantom point, because the differential equations at the other points do not use this point: It can be shown that in the Laplacian, evaluated at ${s}^{(1)}$, the terms using function values at ${s}^{(-1)}$ cancel out. Hence, the problem is closed by the boundary conditions $W^{(0)} = \Phi^{(0)} = 0$.

\bibliographystyle{h-physrev}
\bibliography{literature}
\end{document}